\documentclass{article}
\usepackage{graphicx} % Required for inserting images
\usepackage{amsmath}
\usepackage{subcaption}
\usepackage{appendix}
\usepackage{comment}
\usepackage{setspace}
\usepackage[a4paper, total={6in, 9in}]{geometry}

\title{Complex hypergraph analysis of Australian MPs' professional connections, 1947-2019}

\author{
  Eve Cheng\\
  \texttt{wenjun.cheng@anu.edu.au}
  \and
  Danny Cocks \\
  \texttt{daniel.cocks@gmail.com}
  \and
  Pat Leslie\\
  \texttt{patrick.leslie@anu.edu.au}
       }
\date{\today}

\begin{document}

\maketitle

\begin{abstract}
\noindent We propose a suite of methods to analyse the professional networks of MPs, showing how to analyse weak-tie connections between legislators and the connections between background characteristic attributes. Applied to a novel dataset on the backgrounds of Australian MPs in the Australian Labor Party and the Liberal Party of Australia (1947-2019), we show that our approach can help to describe and explain the decline in working-class and trade unionist MPs from the Labor Party, the homogeneous elitism of the mid-20th century Liberal Party, and the increasing similarity of both parties' professional networks, occurring in the period of party cartellisation from the 1980s onward. Our paper’s findings show that our method has clear potential for broader applications in the study of political representation, diversity, and  elite political networks.
\end{abstract}

\onehalfspacing

\section{Introduction}

Politicians' background data is often used in quantitative political science literature to analyze various political processes, including career success \cite{norris_passages_1997,kam_ministerial_2010, allen_linking_2013, curtin_sex_2022}, substantive representation \cite{best_transformation_2001, barnes_pink-collar_2021}, trends in political recruitment and party organization \cite{katz_changing_1995, miragliotta_legislative_2012, cross_evolving_2014}, and other phenomena like political speech or candidate nomination processes \cite{box_steffensmeier_dynamic_1992, slapin_ideology_2018, dowding_australia_2021}. Tools for the analysis of these data fall into two categories. The most widely used category focuses on what we term `attribute analysis' -- the prevalence of particular backgrounds, either as a descriptive indicator or as an explanatory variable in the analysis of political outcomes. The second understands dyadic networked relationships between politicians as a quantity of interest.

Attribute analysis uses categorical data analysis and regression modelling to show how outcomes differ between politicians with readily available background characteristics such as gender, profession and education. The approach has been instrumental in explaining the descriptive and substantive representation of women, class and race \cite{carnes_why_2016, ogrady_careerists_2019, lowande_descriptive_2019, lotte_hargrave_no_2022} as well as the professionalisation of politics \cite{king_rise_1981, cairney_professionalisation_2007, allen_what_2020} and many other political outcomes. Insofar as they are used for inference, these methods treat politicians as independently and identically distributed units of analysis. Of course, conceptually, the treatment of background data in this manner is a methodological convenience, ignoring the influence of networks in determining outcomes. But politicians'  interdependence is an indispensable component of any holistic explanation of political outcomes.\footnote{At least, politicians' own experience of their networks seems an important part of how they explain their roles in parliament \cite{searing_westminsters_1994}.} Studies show how being centrally positioned in a network might be beneficial by analyzing proxies for that central position. Legislators who have a close family tie with a former politician tend to be more successful in elections and attaining higher office \cite{dal_bo_political_2009,asako_dynastic_2015, smith_political_2017,fiva_political_2018,  smith_dynasties_2020}; politicians with degrees from elite universities are far more likely to be selected to cabinet positions \cite{turner_zwinkels_pathways_2020}. 

Direct analysis of politicians' social or professional networks is far less common. This is likely to be related to the fact that collection for analysis of dyadic data structures has factorial growth; every politician added to the data requires $n-1$ new data points, creating significant barriers for researchers wishing to collect full data sets, particularly in historical settings. There are some exceptions. Studies using last names have pieced together network analysis of close family ties in 19th-century Chilean politics \cite{bro_structure_2020}. Kinship ties have also been shown to affect resource allocation among households in the Philippines \cite{fafchamps_family_2020}. However, for analysis of professional networks, data collection is limited to office-centric networks, treating a particular individual or office as the centre of the network, and tracing professional, familial, contractual and donation connections outward \cite{hill_my_2020}.

To bridge the gap between these two categories of analysis, we introduce a attributes-as-networks method. This approach uses information from politicians' professional networks while collecting only individual-level attribute characteristics as data. Our definition of a network tie differs from a dyadic network. Instead of direct dyadic ties between individuals, we analyse shared attribute characteristics using graph and hypergraph methods. Approaching political networks in this way borrows conceptually from the `strength of weak ties' \cite{granovetter_strength_1973, granovetter_strength_1983}, a theory of network embeddedness that hypothesises that acquaintances (weak tie connections) are more likely to be influential in passing important information from a broader network than a close friend. Recent causal analysis finds that this hypothesis is supported by evidence from LinkedIn recruitment advertisements  \cite{rajkumar_causal_2022}.

In this paper, we analyze the global features of MP and MP background attribute networks to understand their structure and long-term trends in the composition of MP networks for the Liberal Party of Australia (LP) and the Australian Labor Party (ALP). An attributes-as-networks approach promises several advantages over a attribute analysis approach: it allows for  direct structural analysis, incorporating global measures such as network connectivity and directly modelling the dependence between politicians. Instead of assuming that particular types of attributes such as elite education, profession or dynasty membership confer networking advantages, we ask: `How well networked are the attributes themselves?'. In this way, we can examine the composition of weak-tie networks and answer new questions about the mechanisms for networked advantage in politics. 

We also show that we can go a step further to ask `What causes the networks to be well/poorly connected?'. Our approach allows for a good deal of flexibility in representing MP networks to separate underlying structures from noise. Firstly, we constructed graphs for the MPs and hypergraphs for the attributes. MP graphs, while intuitive, often share common attributes, forming complete sub-graphs which in turn create redundancy and noise in the data. Hypergraphs allow us to study the attribute networks to look at the networks from a different perspective  \cite{lung2018hypergraph,barton2022hypergraphs}. In addition to the two sets of graphs/hypergraphs, we can define different edge weights and s-filtration, which allows us to see the role of strong/weak ties/communities in the networks. A second flexibility comes from the choice of graph summary statistics. We introduce two measures of network connectivity: \textit{average maximal flow} and \textit{transitivity}. These measures can both show the level of connectedness at a given time and identify certain network structures that contribute to high/low connectivity. In some cases, a network composition that we call a `bouquet structure' was prevalent in the professional networks of MPs. This structure emerges in conditions where transitivity is higher than expected, while average maximal flow is low. It resembles a bouquet in that two large sub-network groups are separated by a very small number of individuals who connect the two sub-networks together. This sub-network can give a very visual representation of the structure of the networks. When transitivity is low and average maximal flow is high, we sometimes expect some weak structures that consist of `circles' that consists of more than three nodes. These circles can be identified as clusters with weak correlation of higher orders, which again can be visually helpful when analysing network structures. 

The third flexibility comes from the statistical analysis of the networks. We use random graph generation as statistical baseline as comparison. Two sets of random graph generation were generated. The first set assumes the politicians each year have almost absolute freedom to choose their own backgrounds randomly, given the background options that year. The second set, on the other hand, restricts that freedom and assumes the number of MPs in each background shall be fixed. The purpose of the first set of the random graphs is to observe the relative magnitudes of the summary statistics compared to that of the random distribution of MPs and random structure of the networks. The second set focuses on the effects of random structure of the networks only. By simulating networks with different assumptions about the properties of the networks, we can narrow down the sources for the high/low connectivity for each year against a baseline simulation.

In our MP network analysis, we find that the bouquet structure is of particular importance. It typified ALP party organisation in a period where trade unionists with backgrounds in manual labour were routinely selected as candidates. While loosely connected by their union experience, they had little in common with the elite wing of the party whose backgrounds were professional, university-educated and with military experience. In our example from the ALP's network from 1960, the position of chief whip was held by George Lawson, a figure of high network centrality who had a rare combination of military and trade unionist background attributes. This is a structure that scholars of the history of labour movements would recognise. Our analysis of attribute networks shows that the centrality of the trade unionist and manual labourer attributes in the ALP was low from the beginning and decreased over time. These attributes were not connected via education or military service to the rest of the professional network of the ALP and dwindled in number. Conversely, the attribute network of the mid-20th century LP was characterised by higher-than-average connectivity in average maximal flow and connectivity. This implies that the backgrounds of LP MPs of this period were sufficiently interchangeable to be largely immaterial to party organisation. Over time, the distinctiveness in the MP and attribute networks has decreased, as party recruitment has converged on a cartellised recruitment model. Taken together, these findings corroborate and shed new light on studies of party organisation convergence internationally and in Australia \cite{katz_changing_1995, miragliotta_legislative_2012}.

Our paper is arranged as follows, we begin with our application to professional networks among Australian MPs, defining our graph and hypergraph in detail. We then introduce the two summary statistics we use to describe the connectivity of the networks each year. We then describe our two methods for simulating random graphs as statistical analysis. The results are split into two sections, discussing our findings with respect to MP networks and background attribute hypergraphs.\footnote{A brief note on terminology: in this paper we use the terms graphs and networks interchangeably. Our analysis of MP networks follows the usual terminology for network analysis. Our hypergraphs of MP attributes reduce to s-line graphs (discussed in Section \ref{prelim}), which are then analysed using the same measures of connectivity as the MP networks. Our use of the phrase `MP background attributes' or similar refers to particular background characteristics that may (or may not) be shared by other members of an MP's parliamentary party.} We conclude by summarising our argument and suggesting further avenues for study.

\section{Method}
In this section we introduce how we define the networks using graph/hypergraph theory, the summary statistics we use to describe the networks, and the different random graph simulation methods used to provide statistical analysis. These sections serve as three linked but independent sections that show the advantages of using graph/hypergraph theory to study network structures. 

First we define the networks of both MPs and attributes using graphs and hypergraphs. Following the definitions, we introduce the two summary statistics we focus on in this paper: the average maximal flow and the transitivity. We then present the random graph simulations to compare with the real data to provide the statistical analysis for our result. The preliminary background for graph and hypergraph theory is in the appendix.
\subsection{Graphs and hypergraphs definition}
\begin{figure}[h]
    \centering
    \includegraphics[width=\linewidth]{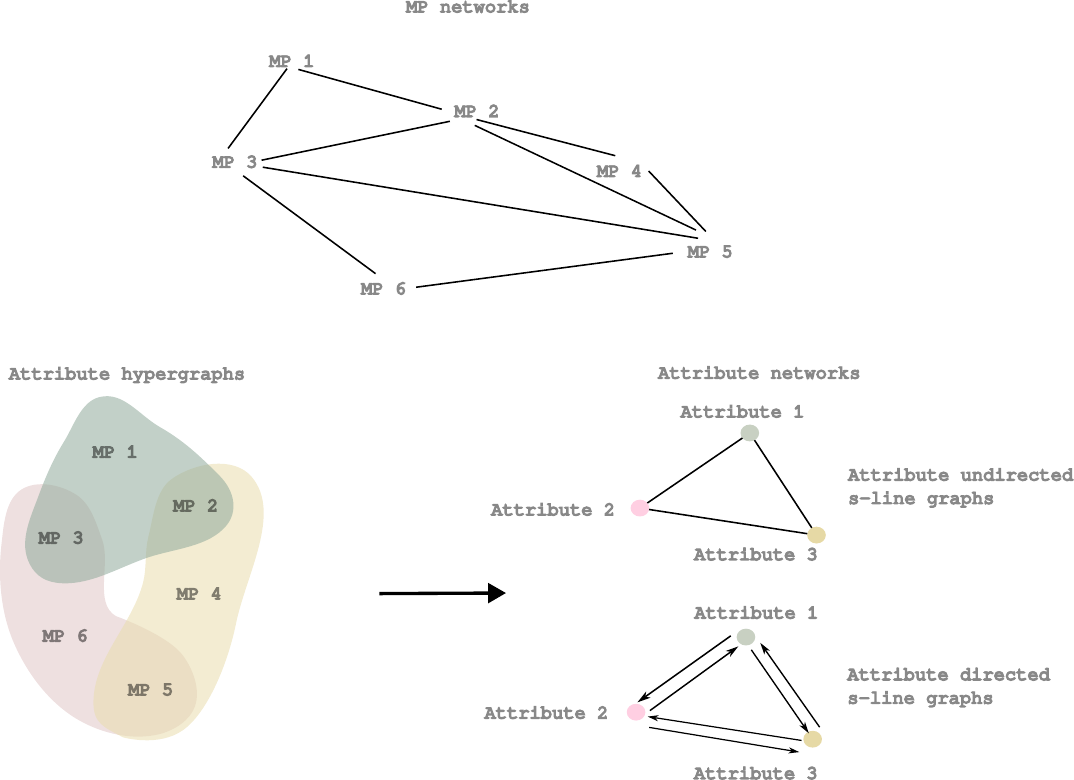}
    \caption{This scheme describes both the graph set-up for MP networks and the hypergraph setup for the attribute networks. Both the MP networks and the attribute hypergraphs have MPs as nodes and shared attributes as edges. The attribute hypergraphs are mapped to two sets of s-line graphs. The first set is undirected s-line graphs, and the second set is directed s-line graphs. }
    \label{setup}
\end{figure}
 For our analysis, we want to capture the important global structural features in both the MP networks and attribute networks (see Figure \ref{setup}). The MP networks uses MPs as vertices and we draw ties among the MPs who shared background attributes such as party, education, military service and prior occupation as edges. How closely the ties are (or \textit{geodesic}) is reflected in the definition of edge weights in the adjacency matrices for the graphs:
\begin{align}
w_{ij} = \sum_{\text{shared attribute k}} \frac{1}{\text{No. MPs with attribute k}}
\end{align}
where $i$, $j$ refer to indices for MPs.

Intuitively, the difference between graphs and hypergraphs is the way we draw ties (or edges) among the entities (or nodes) of the networks. In graphs, we define ties only between pairs of entities. While in hypergraphs, the ties can be drawn among any number of entities. For example, if person A, B, C all share a close family tie, the ties in graphs would be represented using three ties, A-B, A-C, B-C. In hypergraphs, such close family tie is represented by one hyperedge: A-B-C. Because of this feature, hypergraphs are often considered a cleaner, less noisy approach.

While analysing many mathematical structures, it is often beneficial to study the \textit{dual} aspect of said structure in order to gain further insights. If we view the MP networks as collapsed MP hypergraphs which use the career attributes as nodes and MP as hyperedges, we see that the \textit{dual} structure of the MP hypergraphs is the attribute hypergraphs. In our definition of the attribute hypergraphs, we take the MPs as vertices and attributes as hyperedges. In contrast to the MP networks, which record all the pair-wise relationships among MPs, the attribute hypergraphs record the set-wise relationships among the attributes. Because of the set-wise feature, we can further analyse the hypergraphs through a ``flattening" procedure and reduce them to s-line graphs.

The ``flattening" procedure is where the hypergraph approach cleans up the noise and produces our attribute networks. In MP networks, each attribute, for example, being an alumni of University of Sydney, causes a great blob of all MPs who went to University of Sydney. In the ``flattened" attribute networks, each great blob is reduced to a single node of University of Sydney. The result from this ``flattening" procedure are ordinary graphs. But we call them line-graphs to remind ourselves of their hypergraph origin. The \textit{s} in s-line graphs corresponds to the filtering of the ties among the attributes. Because the intersection between hyperedges can contain arbitrary number of MPs, we can filter the line-graphs based on the strength of the ties. The \textit{s} corresponds to the minimum number of MPs who live in the intersection before the tie is acknowledged in the attribute networks. Further technical details see Appendix A.

The networks are named for the feature which forms the vertices of the graph: the MP network focuses on MPs as vertices, whereas the attribute network, when converted to an s-line graph, represents attributes as vertices of a graph.

Other than noise deduction, hypergraphs also introduce diversity into the definition of the strength of ties. There are two sets of s-line graphs to which we map the attribute networks (see Figure \ref{setup}). For the first set of s-line graphs, which we call undirected s-line graphs, we do not define any direction between the ties. The ties between, for example, being an alumni of the Australian National University and holding the office of Prime Minister, only reflect how strongly the correlation is with no orientation. Additionally, we can choose whether to reflect the heterogeneous strengths of the ties in this set of s-line graphs. So the first set as undirected s-line graphs can be further divided into weighted or unweighted graphs. For unweighted undirected s-line graphs, we set all edges of the s-line graphs to be uniformly 1. 

For weighted undirected s-line graphs, we want to reflect a \textit{normalised} strength of the ties through weights of the edges. The weights of the edges are calculated from the parent hypergraph the line-graph originates from:
 \begin{align}
  w_{i,j} = \frac{\text{no. MPs in } e_{i} \cap e_{j}}{\text{no. MPs in } e_{i} \bigcup e_{j}}
 \end{align}
 where $i$ and $j$ are indices for the attributes, $e_i, e_j$ represent the two hyperedges corresponding to two attributes. This is equivalent to the Jaccard index for the pair-wise sets of hyperedges. We normalise the edge weight with the sum of the size of the hyperedges to ensure that the edge weight reflects the proportion of the overlap between the hyperedges and the size of the two hyperedges.

The second set of s-line graphs are directed s-line graphs, in which we split each edge of the undirected graphs into two edges, one for each direction. We introduce the extra direction on the ties between attributes in hope to capture a sense of \textit{flow} between the attributes: how likely it is for someone to be in the military given their university background? Similar to the transition rate in statistical physics, instead of defining a single edge between a pair of nodes, we distinguish between the edge $e_{i,j}$ and $e_{j,i}$. The weights in the directed s-line graphs also reflect this asymmetry in the edges:
 \begin{align}
  w_{i,j} = \frac{\text{no. MPs in } e_{i} \cap e_{j}}{\text{no. MPs in } e_{i} }
 \end{align}  

Here we can see that the diversity of graph/hypergraph definitions provides the flexibility for investigating different properties of the networks easily and directly. This will be further explored in the result section.

We calculate the summary statistics for graphs and line-graphs with different edge weights. Since there is only one edge weight definition for the MP networks, the analysis concerning different edge weights is only relevant for the attribute networks. The different edge weights can show how the strength of the connections among hyperedges affects the overall networks. Intuitively, the unweighted line graphs report lower/higher connectivity than the weighted ones if there are fewer/more weakly connected edges. If the weighted undirected and directed both exhibit higher/lower connectivity compared to the unweighted option, we will have good reasons to believe that it is due to the amount of weakly/strongly connected edges in general. But what about when the different weighted options disagree with one another? 

To interpret the differences between these two weighted options, we have two predictions (see Figure \ref{substruct}(c)). The first prediction is that the directed weighted option is more sensitive to the asymmetry of the edges when one of the edges is very small, but less sensitive if one of the edges becomes very large, assuming the other edge is average. In other words, with $\text{no. nodes in } e_{i} \cap e_{j}$ (see Section \ref{MCdef}) fixed if we reduce/increase the size of $e_{i}$ to create more asymmetry between $e_{i}$ and $e_{j}$, the relative changes reflected in the average maximal flow in the directed networks will be larger/smaller than that in the undirected one. 

The second prediction is that the increase in the overlap ($\text{no. nodes in } e_{i} \cap e_{j}$) is reflected more in the average maximal flow in the undirected weighted networks compared to the directed networks (more details in the appendix).  

Finally, the s-filtration preserves the highly connected part of the networks and filters out the weak edges (definitions see Appendix A). This has a similar effect to the weighted edges. However, since the filtration weakens the contribution from the weak edges by deleting them completely, the result is expected to be more dramatic than the edge weights. By filtering out the weak edges, we believe that the filtering can reveal the more important, close-knit structures in the networks. 

\subsection{Summary statistics: measures of connectivity \label{MCdef}}
In this paper, we focus on global features that characterise the connectivity of both MP and attribute networks defined above. The measures we introduce are the average maximal flow and the transitivity. While both allow for summary statistics of network connectivity, our analysis shows that they measure different aspects of professional networks and complement each other in describing the politics of professional networks.

The average maximum flow is used to study the changes in the structure of the graphs over the years. It can be calculated as follows:
\begin{align}
\text{averF} = \frac{\sum_{\{(i,j)| i,j \in V, i \neq j\}} F(i,j)}{|\{(i,j)| i,j \in V, i \neq j\}|}
\end{align}
where V is the set of all vertices in the network, F takes any two nodes and returns the maximal flow between them. The edge weights defined above are passed as the capacity for each edge in E in the form of a map $c: E \rightarrow \mathcal{R}^+$, which reflects the maximum flow that each edge can allow \cite{edmonds1972theoretical}.

Transitivity is a topological parameter that gives information on how clustered the networks are \cite{wasserman1994social,estrada2005complex}. It is defined as:
\begin{align}
T(G) = \frac{3 \times \text{number of triangles}}{\text{number of connected triples}}
\end{align}
It can be loosely understood as the probability of A being related to C, knowing A is related to B and B is related to C. The higher the value, the more closely knit the community is.

The two measures of connectivity, average maximal flow and transitivity, are simple to interpret individually: the larger the measure, the more connected the network is. However, we would like to know more about the kind of connectivity present. Unfortunately, the two measures are not directly comparable numerically and may have different dependence on graph size. By analysing simulated networks, which are discussed below, we can directly compare the relative connectivity for the two measures. In particular, when we say one measure suggests higher/lower connectivity than the other in this section, we mean that the relative difference between real and simulated is higher in one measure than the other. Because the ways the two connectivity measures are calculated is different, they can disagree with one another. The disagreements can be  interesting as they inform us about the structures of professional networks among MPs. 
\begin{figure}
\centering
 \begin{subfigure}{0.13\textwidth}
     \includegraphics[width=\textwidth]{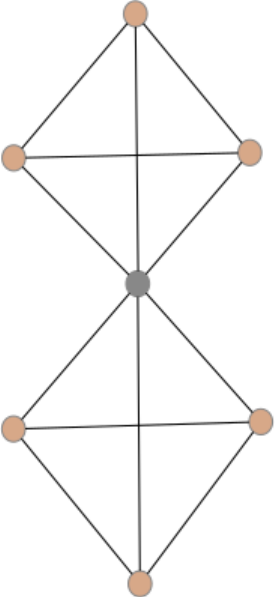}
     \caption{}
  \end{subfigure}
\hfill
 \begin{subfigure}{0.32\textwidth}
     \includegraphics[width=\textwidth]{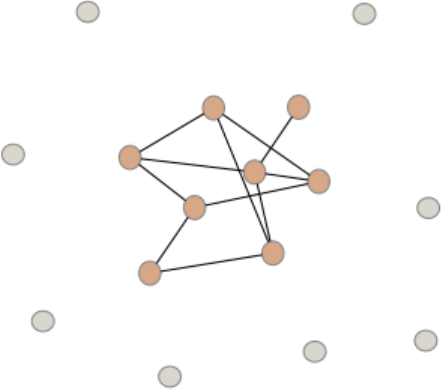}
     \caption{}
  \end{subfigure} 
\hfill
 \begin{subfigure}{0.32\textwidth}
     \includegraphics[width=\textwidth]{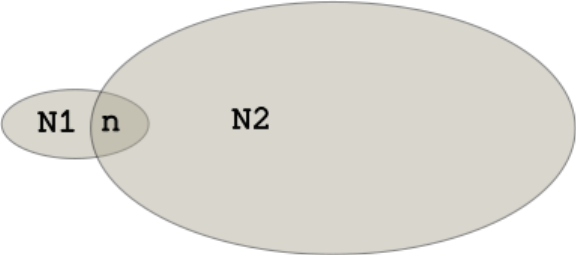}
     \caption{}
  \end{subfigure} 
\caption{(a) an example of a bouquet structure, with one node connecting two complete networks. (b) an example of a satellite structure, with a compact structure in the middle and scattered disconnected components. (c) asymmetrical bonding between a large edge and a small edge, where N1 and N2 represent the number of nodes the edges possess and $N1 << N2$ and n is the number of nodes that are in the overlap region.  }
\label{substruct}
\end{figure}
 When the transitivity suggests much higher connectivity than average maximal flow, we observe two sub-structures that appear to explain this anomaly (see Fig \ref{substruct}(a)-(b)). The first is the bouquet structure, which consists of a connecting small structure (in the limiting case a node) and several connected components being linked by it. In our data, a bouquet structure implies that the professional networks of MPs are divided among two or more densely networked groups. The average maximal flow is very sensitive to this structure since the average maximal flow can decrease dramatically in the bouquet structure, while the transitivity can be largely unaffected. The details on the justification for this claim can be found in the appendix C. Another sub-structure that can contribute to this phenomenon is the satellite structure (see Fig \ref{substruct}(b)). In this case, since transitivity does not take into account the number of unconnected nodes, it is not reduced by an increased number of satellites, while the average maximal flow is greatly affected.

Finally, since transitivity focuses mainly on the connected triples while maximal flow considers larger structures, it is also common to have unimpressive transitivity while maximal flow suggests high connectivity. It just means that the structure is highly connected but it is not always possible to go directly from one node to another. For example, structure such as circles of nodes that are made up of more than three nodes, like squares, pentagons, or more can potentially lead to this observation.

\subsection{Random graph simulations}
% \begin{figure}[t]
%     \centering
%     \includegraphics[width=0.8\linewidth]{images/drawing-1.pdf}
%     \caption{This diagram demonstrates the relationship between the real data and the simulation data. The pink boxes on the top correspond to the real attributes such as Unimelb, ANU and UQ, and the dots within the boxes represent the number of MPs from these universities. The simulation boxes are the green ones down below. The simulation randomly shuffles the MPs to different boxes and generate a random network based on the number of MPs and the attributes.}
%     \label{simulation}
% \end{figure}
In this section, we introduce the statistical analysis of our result using random graph simulations. We do these simulations for two reasons. First, we wish to take into account the particular dependency of the average maximal flow measure on the size of the graphs, the number of attributes present each year and the number of unknown present in the historical records. The size dependency can be observed in a simple thought experiment of complete graphs of size n, whose average maximal flow can be shown to be n-1, which is linearly dependent on n. Second, by using these simulations to ``fit" the real summary statistics, we find out the defining characteristics of the networks. If we encounter great difficulty in fitting the real statistics, we can then explore any additional higher dimensional topology that are entirely missed by conventional means \cite{reimann2017cliques}.

We call the first set of simulations \textit{free choice} simulations. In this set of simulations, all MPs are ``free" to randomly choose their career backgrounds from those available each year. More precisely, for each year, 100 random networks are generated for LP and ALP respectively. Each random network preserves the number of MPs, the attributes, the ratio of MPs with military backgrounds and unknowns present each year but allows their relative proportions to vary. This is equivalent to taking the same number of MPs each year and randomly shuffling them into different attributes while the number of attributes and the ratio of military attendance remain roughly the same. The reason we kept the ratio of military attendance roughly the same is because we observed that the military background is overwhelmingly influential, especially in early years. With that influence in mind, we aim to observe the additional influence coming from other attributes. This is an important set of simulations as they tell us how much the real data deviates from what we expect given the available options in terms of career in that year. 

We call the second set of simulations \textit{shuffle}. Faithful to the name, the simulations preserves the number of edges in the networks and merely shuffles them around. This set of simulations focus on the additional effects from the distribution of nodes among the hyperedges, and it poses the question how much of the deviations we observe from the first set of simulations were contributed by the inequality in the number of MPs in each attribute? 

\section{Data}
The data consists of the career backgrounds of the parliamentary members from 1947 to 2019. The data contains the career background information for all the MPs each year including their unique identifier, party, university background, subject at university, their principal occupation before their election, and military experience. Our data is sourced from the Australian Parliamentary Library's Parliamentary Handbook, compiled by the Authors.\footnote{We are happy to share with interested researchers. A skeleton structure for the data is available as an R package at www.github.com/palesl/ausPH.} Our graph structure information is stored in graph database software developed by the authors for easy manipulation of graph and hypergraph structures.

\begin{comment}

There will be another paper following this that analyse the more local features. There are two sets of measures used. The first set of measures contains mostly characteristics of nodes and are presented as either the average for all nodes or descriptors of their distributions for all nodes. They contain the degree and centrality distribution. They serve as an auxiliary tool for the second set of measures.  The second set of measures is what we focus on since it reflects a lot more structural information. 

\subsubsection{Degree and centrality distribution of the networks}
The concept degree is defined in section \ref{prelim}. We also use the centrality distribution. There are many flavours of centrality measures, the one we implemented is betweenness centrality. The definition for the betweenness centrality is as follows:
\begin{align}
c_B(v) = \sum_{s,t \in V} \frac{\sigma(s,t|v)}{\sigma(s,t)}
\end{align}
where V is the set of nodes, $\sigma(s,t)$ is the total number of shortest paths from s to t, and $\sigma(s,t|v)$ is the number of those paths that contains v while v is not s nor t. If v = s or v = t or s = t, the centrality is defined as 0 \cite{brandes2008variants}. The weighted undirected networks were used for calculations for centrality, but the path lengths are defined as the inverse of the edge weights. 

\end{comment}

\section{Results}
\subsection{MP networks}
\begin{figure}
\centering
 \begin{subfigure}{0.49\textwidth}
     \includegraphics[width=\textwidth]{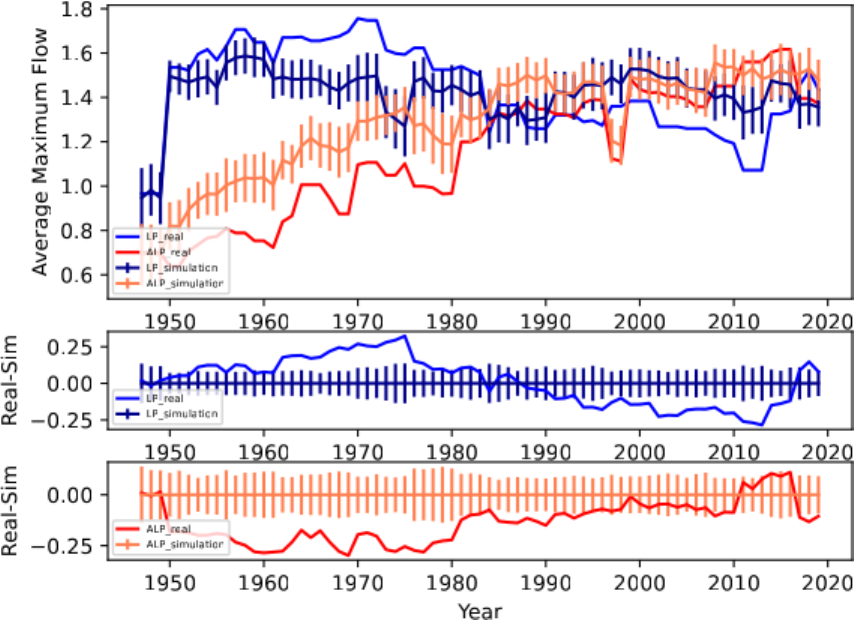}
     \caption{\textit{free choice; average flow}}
  \end{subfigure}
\hfill
 \begin{subfigure}{0.49\textwidth}
     \includegraphics[width=\textwidth]{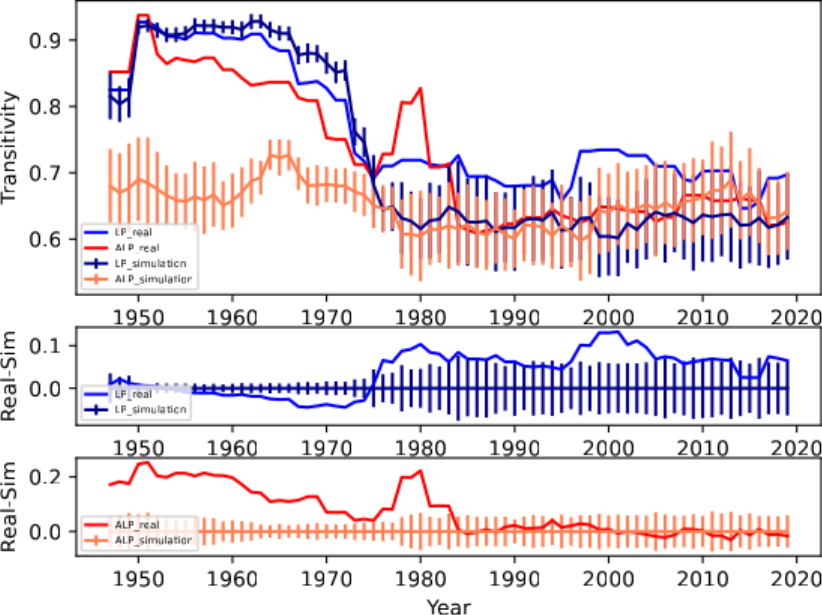}
     \caption{\textit{free choice; transitivity}}
  \end{subfigure} 
  \hfill
 \begin{subfigure}{0.49\textwidth}
     \includegraphics[width=\textwidth]{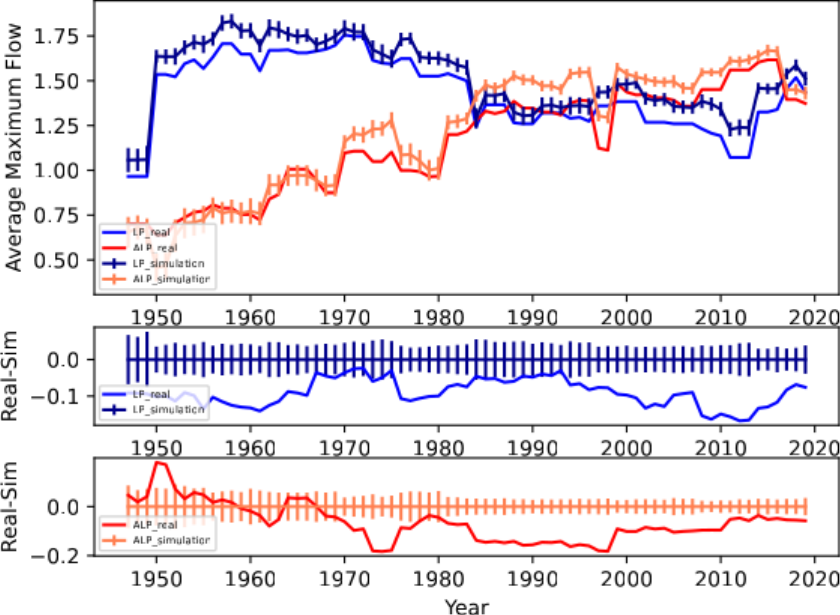}
     \caption{\textit{shuffle; average flow}}
  \end{subfigure}
\hfill
 \begin{subfigure}{0.49\textwidth}
     \includegraphics[width=\textwidth]{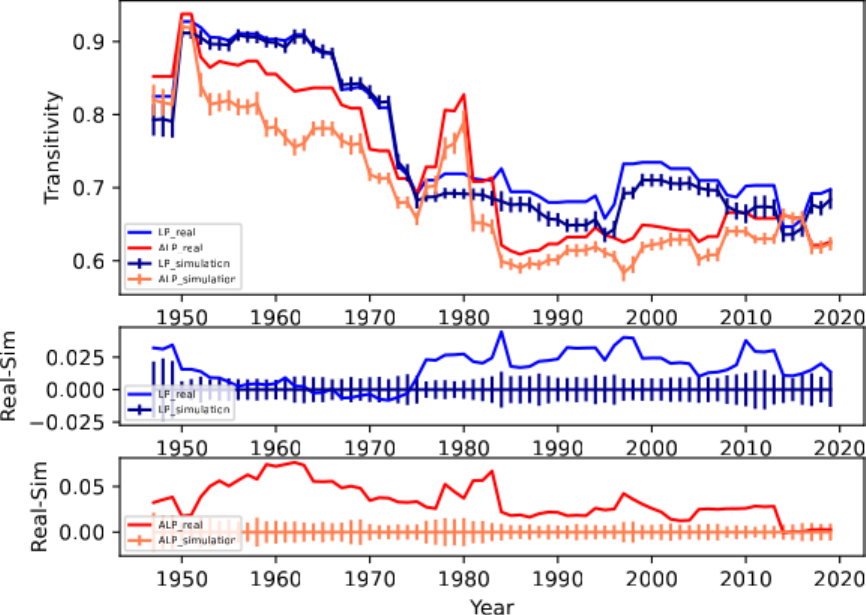}
     \caption{\textit{shuffle; transitivity}}
  \end{subfigure} 
  \hfill
 \begin{subfigure}{0.55\textwidth}
     \includegraphics[width=\textwidth]{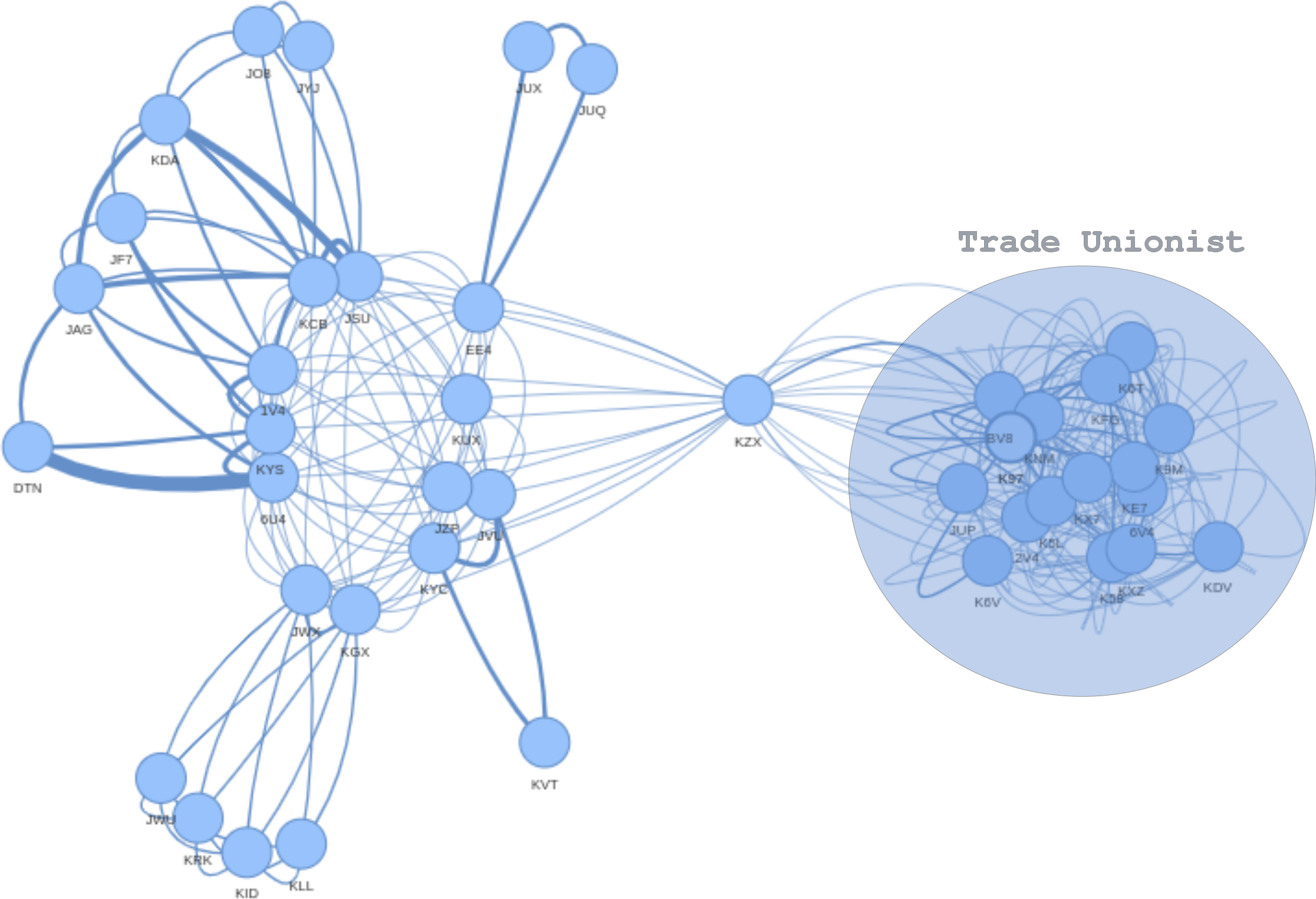}
     \caption{}
  \end{subfigure} 
   \hfill
 \begin{subfigure}{0.42\textwidth}
     \includegraphics[width=\textwidth]{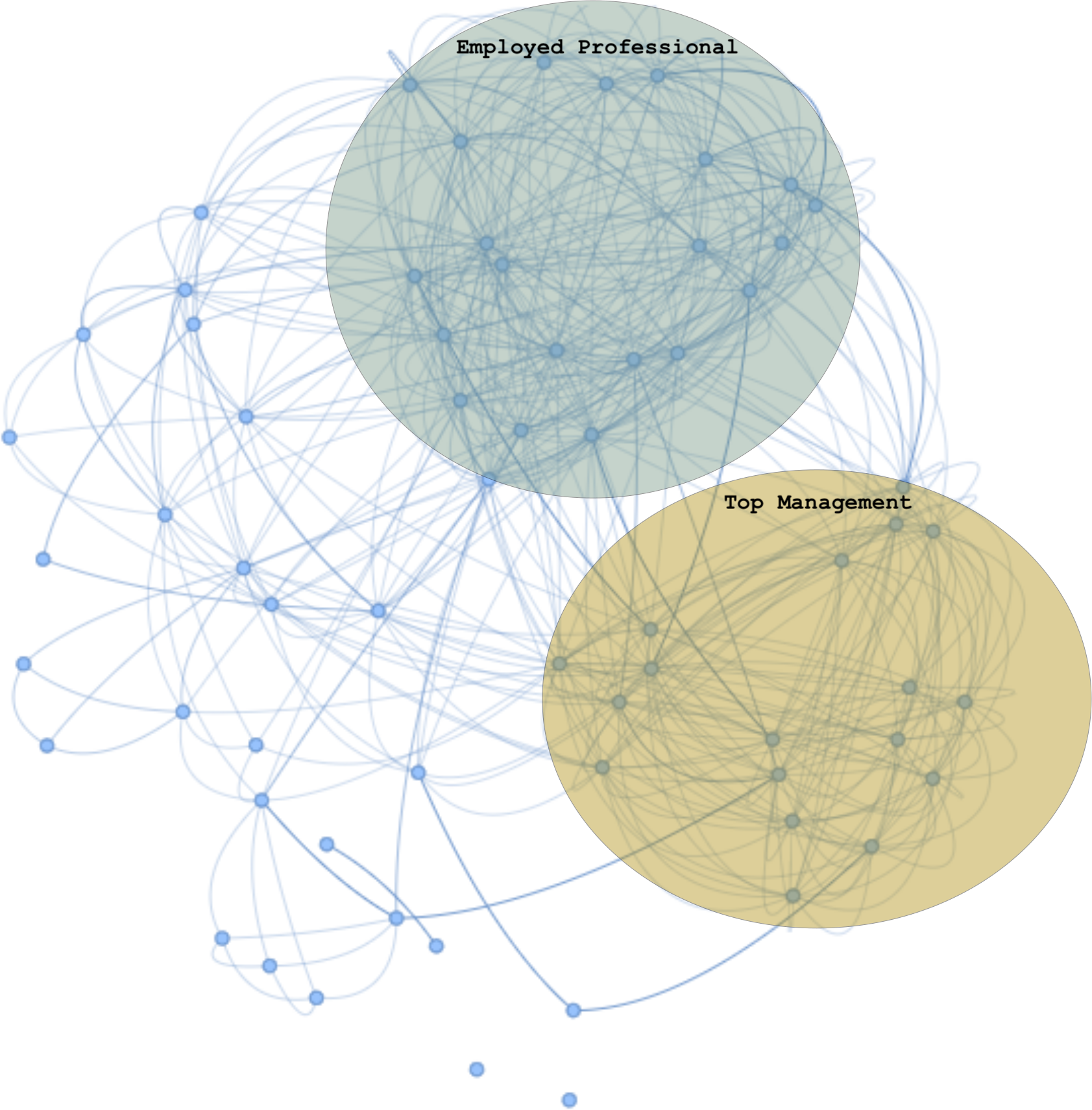}
     \caption{}
  \end{subfigure}  
\caption{(a) and (b) show the comparison between the simulation (\textit{free choice}) and real MP networks for average maximal flow and transitivity respectively. The error bars on the simulation data are the standard deviations of the 100 simulations calculated for each year, with the simulation line being the average of the 100 simulations. (c) and (d) show the same comparison with (a) and (b) except against the \textit{shuffle} simulations. (e) the MP network for ALP in 1960. The bouquet structure is obvious with one end being the trade unionists. (f) the slight clustering structure caused by many bouquet sub-structures (the LP MP network at 2003).}
\label{MPs}
\end{figure}

In the MP networks, MPs are the nodes and the edges represent shared attributes between the pairs of MPs. Figure \ref{MPs} shows both the average maximal flow and transitivity of the MP networks compared to the both \textit{free choice} and \textit{shuffle} simulation networks for each major party over time. Comparing the simulated average maximal flow and transitivity, we observe that the simulated average maximal flow from both methods overall rise while an opposite trend is observed in transitivity. This is expected as the average maximal flow has a dependency on the network size, and the number of MPs increases over the years. On the other hand, transitivity is not affected much by the size of the networks. The opposite trend in transitivity is not surprising as the MP networks are expected to be less connected over time because of the increase in the number of attributes over time (see Figure \ref{number}). More career choices for the MPs lead to less connected networks. 

The more complex structural information of the real networks beyond simple descriptors such as size and number of career choices lies in the deviations between the real network and the simulations. According to Figure \ref{MPs}(a), the major difference between real data and the \textit{free choice} simulations for the Labor Party in average maximal flow occurs before 1980. The real average maximal flow seems significantly lower than the simulations. Interestingly, the transitivity was much higher than simulations before 1980. 

This is, we believe, a result of two things. The first is that we observed many bouquet substructures before 1980. The bouquet substructure contributes to high transitivity and low maximal flow because of the flow-limiting centre. This observation is also supported by the high skewness of the degree distribution in Figure \ref{skew}. This suggests that the ALP at this time was divided between groups of densely connected MPs, connected by a small number of `go-between' MPs with high centrality. In 1960, the central figure (ID: KZX) was George Lawson, the only MP with both experience in the military and the trade union movement. Lawson appears to have been notable for his ability to act as a link between the two branches of the party's professional networks, serving in the cabinet as Chief Whip. 

Secondly, before the 1980s, there were high percentages of trade unionists among MPs without university backgrounds. This leads to low average maximal flow because it means that the trade unionists, who made up approximately 40 per cent of the parliamentary party, compose one weakly connected complete subgraph (see Figure \ref{MPs}(c)). This effect is lessened in the comparison with \textit{shuffle} simulation. By fixing the number of MPs as trade unionists, we observe that the low average maximal flow has been elevated to the simulation baseline. However, this agreement in average maximal flow between \textit{shuffle} simulation and real data does not mean the distribution of MPs in their career backgrounds can fully reproduce the networks. This is evident in the comparison between \textit{shuffle} simulation and real data in transitivity. The persistent high transitivity before 1980s shows the importance of the bouquet structures as previously mentioned.

For the Liberal Party, the most prominent deviations in average maximal flow against \textit{free choice} simulation are 1960-1975 and again from 1990-2015, with the first period possessing higher than average and the second phase lower than the average. In contrast to the figure for comparison against \textit{free choice} transitivity, the first period  (1960-1975) is only slightly lower than the simulation average and the second period (1990-2015) higher than average. 

In terms of the first phase (1960-1975), while the \textit{free choice} simulation suggests higher connectivity of LP MP networks, the \textit{shuffle} simulation shows very little discrepancy. It shows that the high connectivity shown against average maximal flow from \textit{free choice} simulation most likely comes from the distribution of MPs, a large proportion of which are employed professionals (lawyers, managers). Unlike the trade unionist trait in Labor Party, this largely shared trait did not reduce average maximal flow. This is because unlike trade unionists, who are relatively poorly connected to the rest of the party via other attributes such as education and military service, MPs with professional backgrounds share many common traits with others. Once the high number of MPs who were employed professional was taken into account, little difference was observed between the simulation and the real data.  

In the second phase (1990-2015), we observe weak clustering behaviour caused by many bouquet substructures, similar to the structure in the ALP prior to 1980, explaining why the average maximal flow is low, while the transitivity is high (see Figure \ref{MPs}(d)). This difference is lessened in the \textit{shuffle} simulations but still visible. In this case, we find that the reason for the formation of bouquet structures was due to a relatively recent influx of MPs whose prior occupations were not as employed professionals but as senior management within large organisations. Two prime ministers exemplify this generational shift in candidate selection within the Liberal Party: John Howard and Tony Abbott. While Howard had worked as a solicitor for 12 years before entering parliament \cite{howard_lazarus_2010}, Abbott had worked as a journalist and manager of a concrete plant. Most consequentially, he had been director of Australians for Constitutional Monarchy, the group that would lead the `Vote No' campaign to victory in the 1999 Republic Referendum.

\subsection{Attribute networks}
\begin{figure}
  \begin{subfigure}{0.3\textwidth}
  \centering
   \textit{Uniform weights}
     \includegraphics[width=\textwidth]{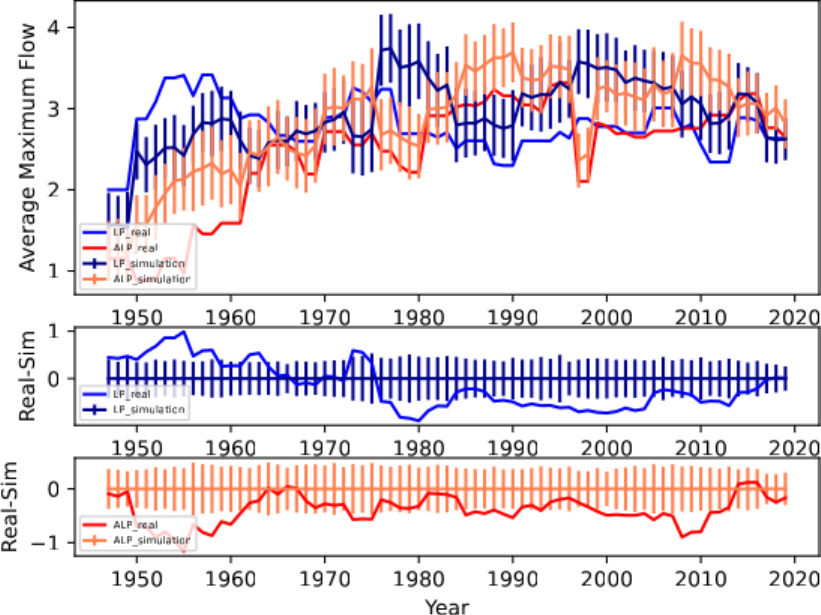}
     \caption{s=0, \textit{free choice}}
  \end{subfigure}
\hfill
    \begin{subfigure}{0.3\textwidth}
     \centering
   \textit{Normalised weights}
     \includegraphics[width=\textwidth]{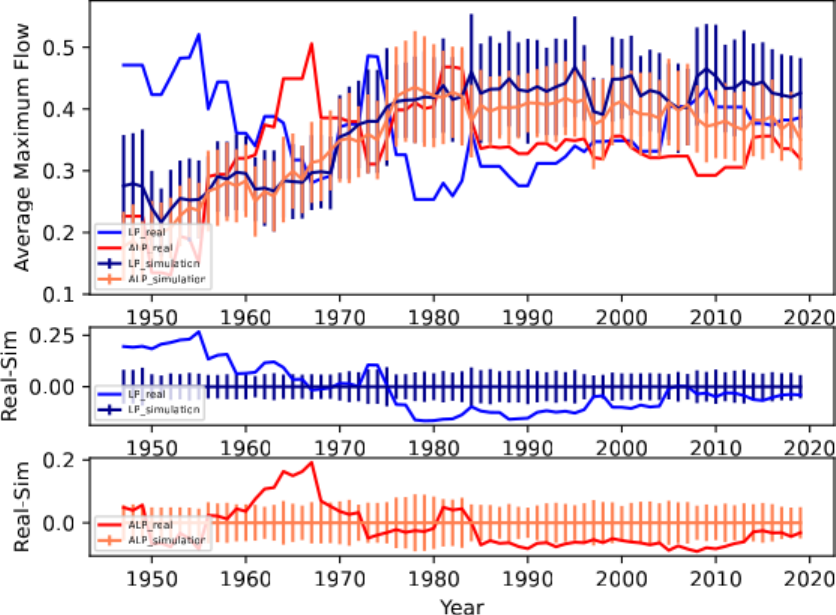}
     \caption{s=0, \textit{free choice}}
  \end{subfigure}
\hfill
\begin{subfigure}{0.3\textwidth}
 \centering
   \textit{Directed normalised weights}
     \includegraphics[width=\textwidth]{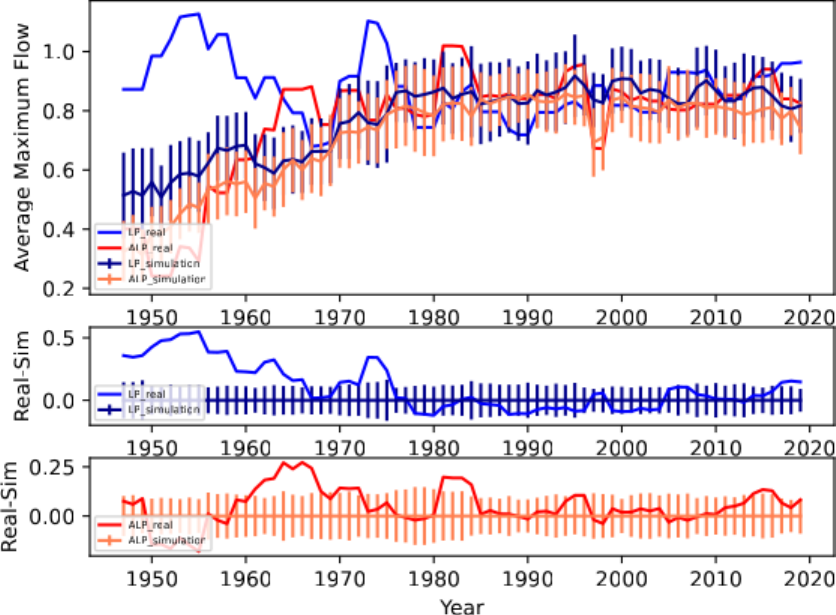}
     \caption{s=0, \textit{free choice}}
\end{subfigure}
\hfill
  \begin{subfigure}{0.3\textwidth}
     \includegraphics[width=\textwidth]{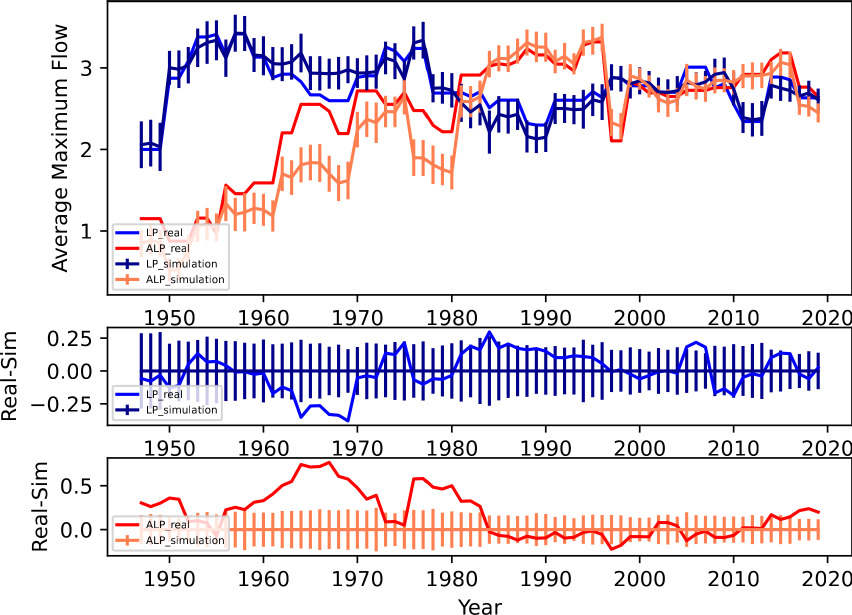}
     \caption{s=0, \textit{shuffle}}
  \end{subfigure}
\hfill
    \begin{subfigure}{0.3\textwidth}
     \includegraphics[width=\textwidth]{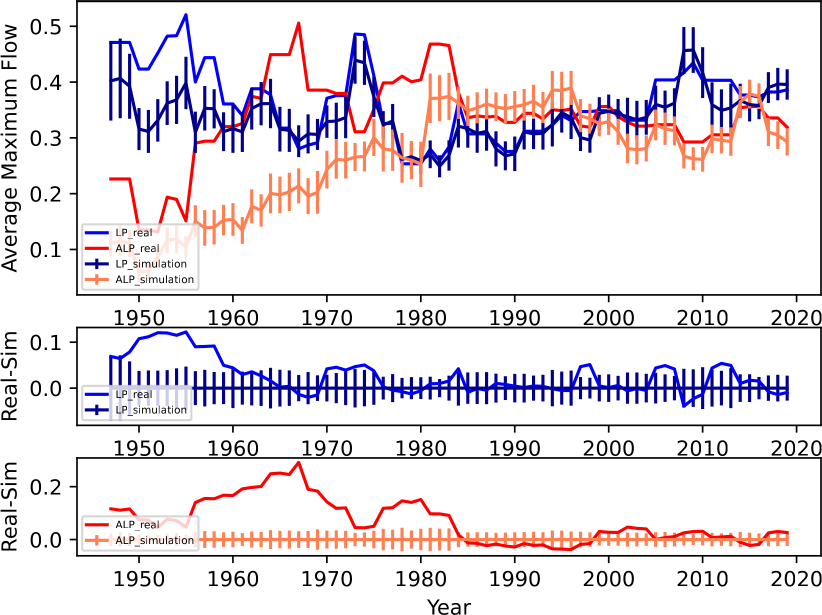}
     \caption{s=0, \textit{shuffle}}
  \end{subfigure}
\hfill
\begin{subfigure}{0.3\textwidth}
     \includegraphics[width=\textwidth]{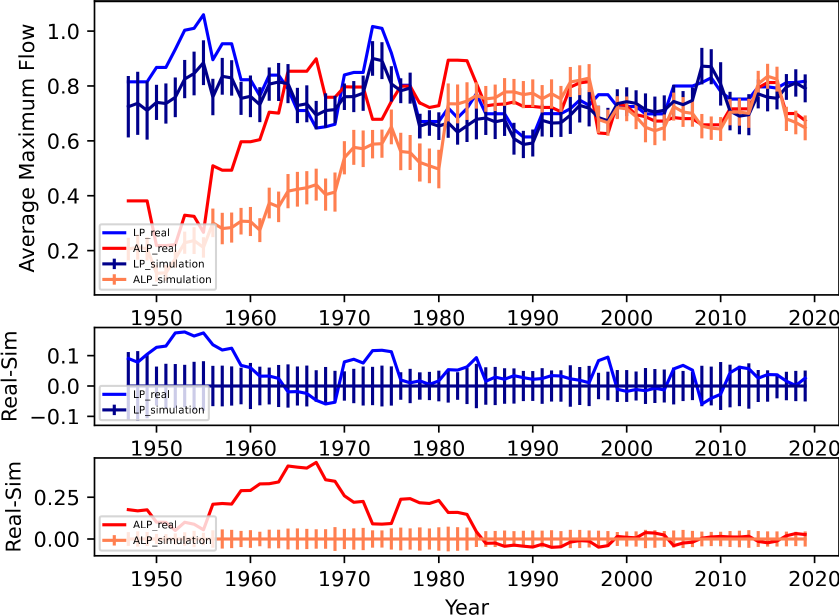}
     \caption{s=0, \textit{shuffle}}
\end{subfigure}
\hfill
 \begin{subfigure}{0.3\textwidth}
     \includegraphics[width=\textwidth]{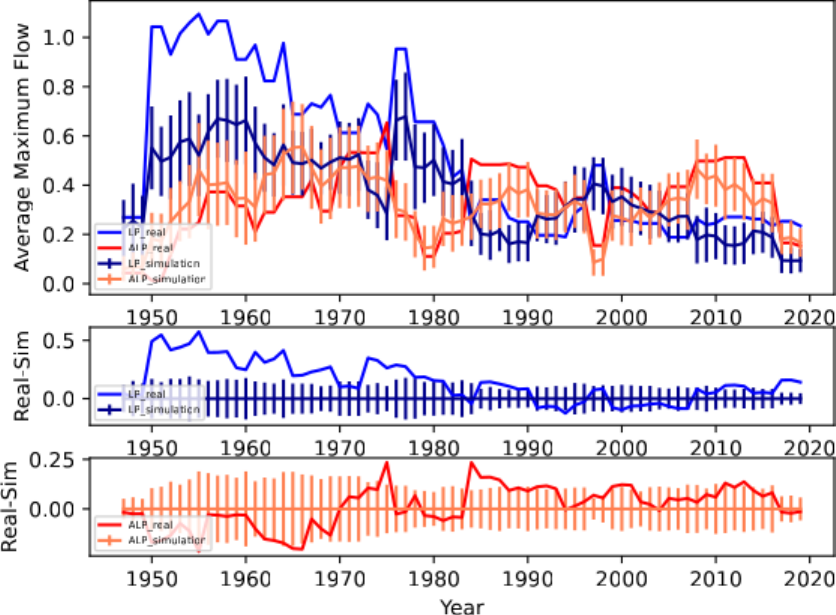}
     \caption{s=2, \textit{free choice}}
  \end{subfigure}
\hfill
 \begin{subfigure}{0.3\textwidth}
     \includegraphics[width=\textwidth]{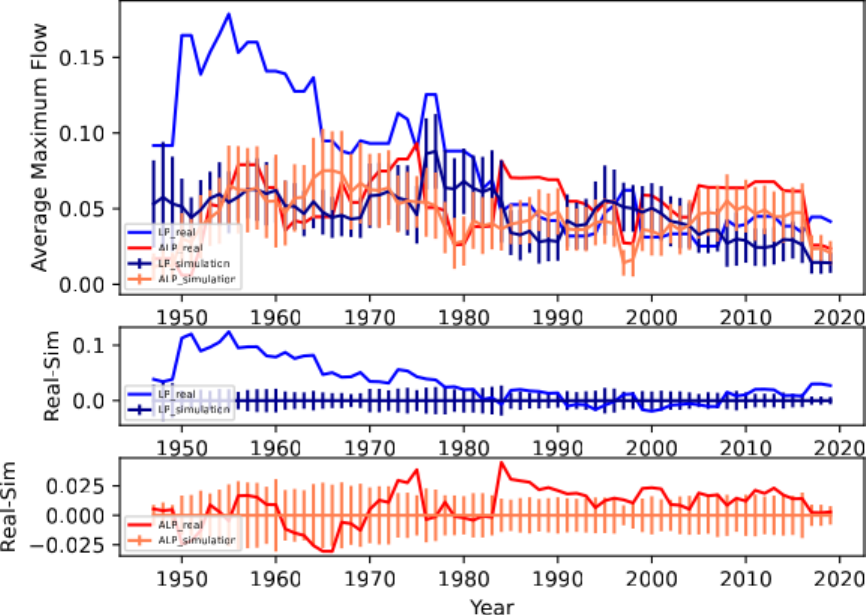}
     \caption{s=2, \textit{free choice}}
  \end{subfigure}
  \hfill
\begin{subfigure}{0.3\textwidth}
     \includegraphics[width=\textwidth]{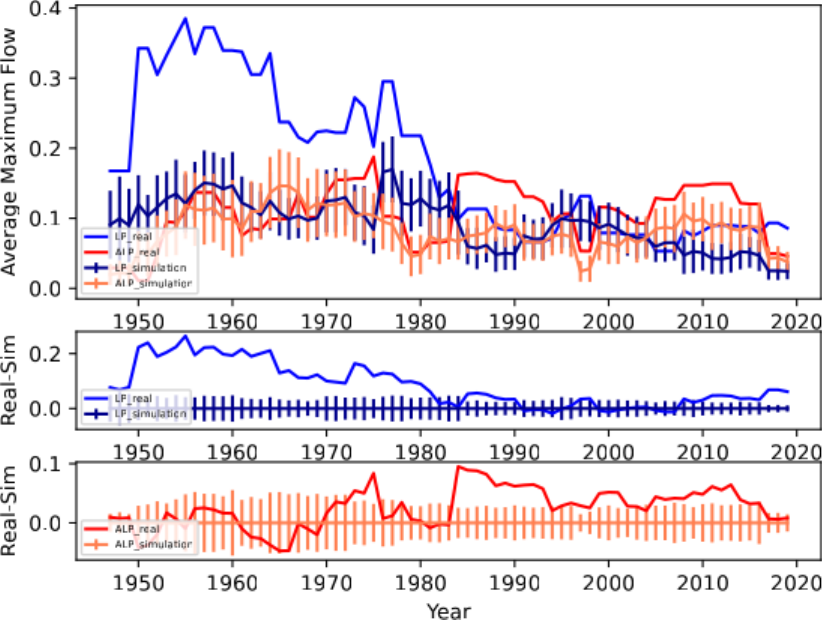}
     \caption{s=2, \textit{free choice}}
  \end{subfigure}
  \hfill
 \begin{subfigure}{0.3\textwidth}
     \includegraphics[width=\textwidth]{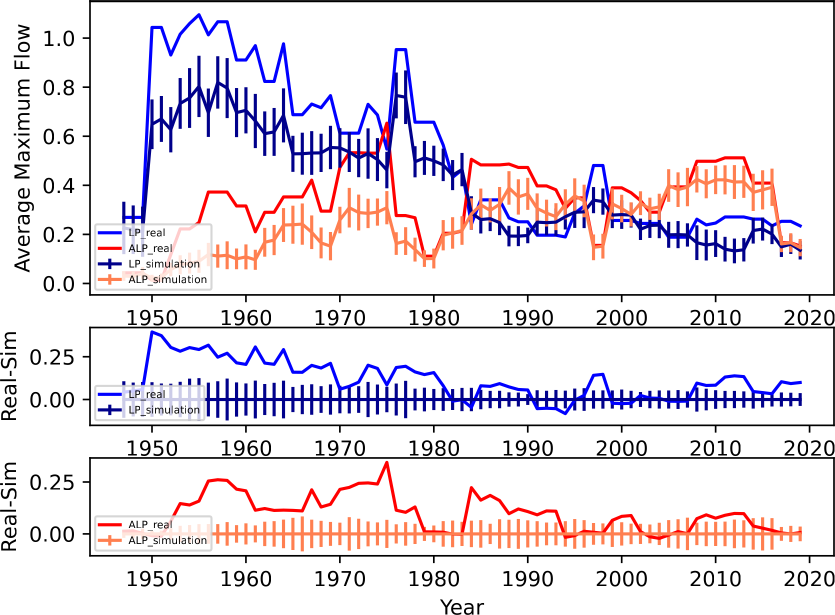}
     \caption{s=2, \textit{shuffle}}
  \end{subfigure}
\hfill
 \begin{subfigure}{0.3\textwidth}
     \includegraphics[width=\textwidth]{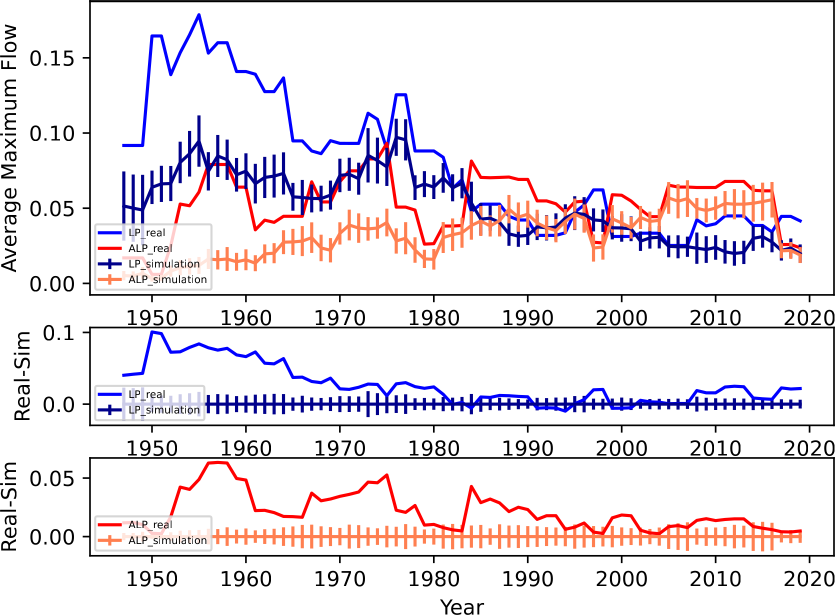}
     \caption{s=2, \textit{shuffle}}
  \end{subfigure}
  \hfill
\begin{subfigure}{0.3\textwidth}
     \includegraphics[width=\textwidth]{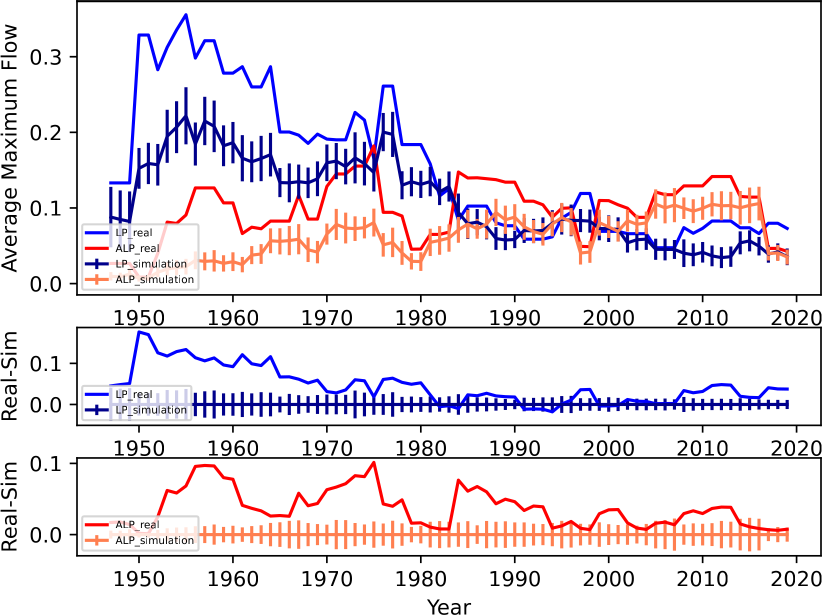}
     \caption{s=2, \textit{shuffle}}
  \end{subfigure}
  \caption{The diagrams (a)-(c) and (g)-(i) show the comparison between simulation and real attribute networks in terms of changes of average maximum flow over the years for both Liberal and Labor parties for \textit{free choice} simulation, s=0,2, and edge weight being uniform, normalised and directed normalised weight from left to right. The diagrams (d) - (f) and (j) - (l) show the similar set of comparisons with \textit{shuffle} simulation. The error bars on the simulation data are the standard deviations of the 100 simulations calculated for each year, with the simulation line being the average of the 100 simulations. }
  \label{classamf}
\end{figure}

\begin{figure}
  \begin{subfigure}{0.45\textwidth}
     \includegraphics[width=\textwidth]{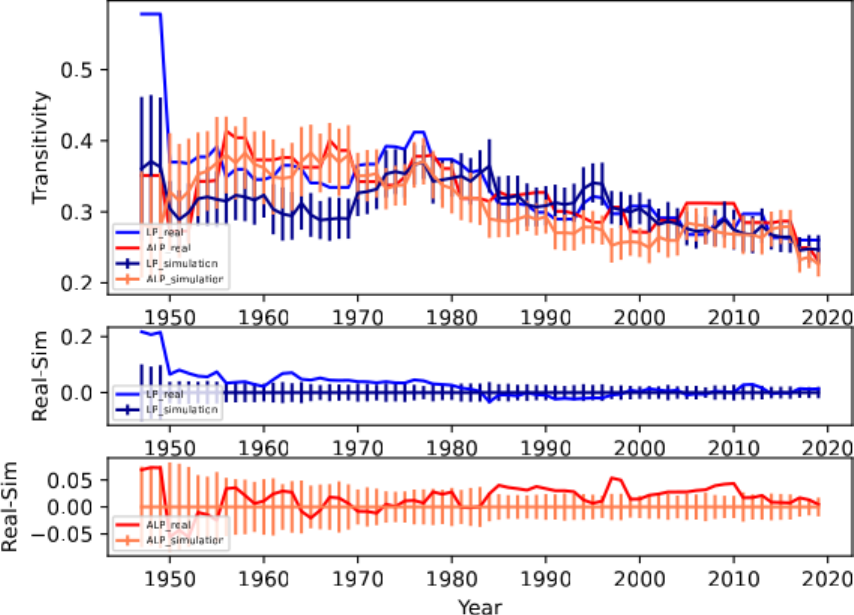}
     \caption{s=0, \textit{free choice}}
  \end{subfigure}
\hfill
  \begin{subfigure}{0.45\textwidth}
     \includegraphics[width=\textwidth]{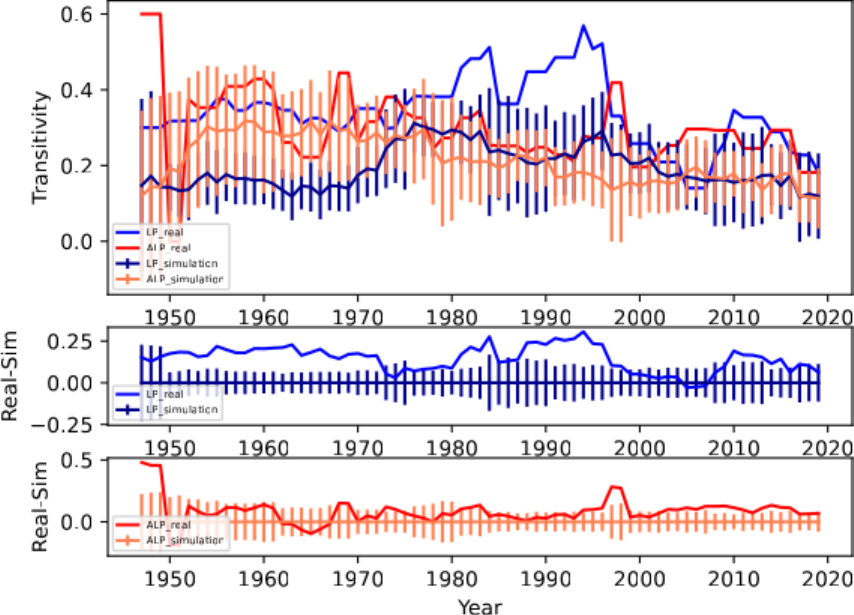}
     \caption{s=2, \textit{free choice}}
  \end{subfigure}
    \begin{subfigure}{0.45\textwidth}
     \includegraphics[width=\textwidth]{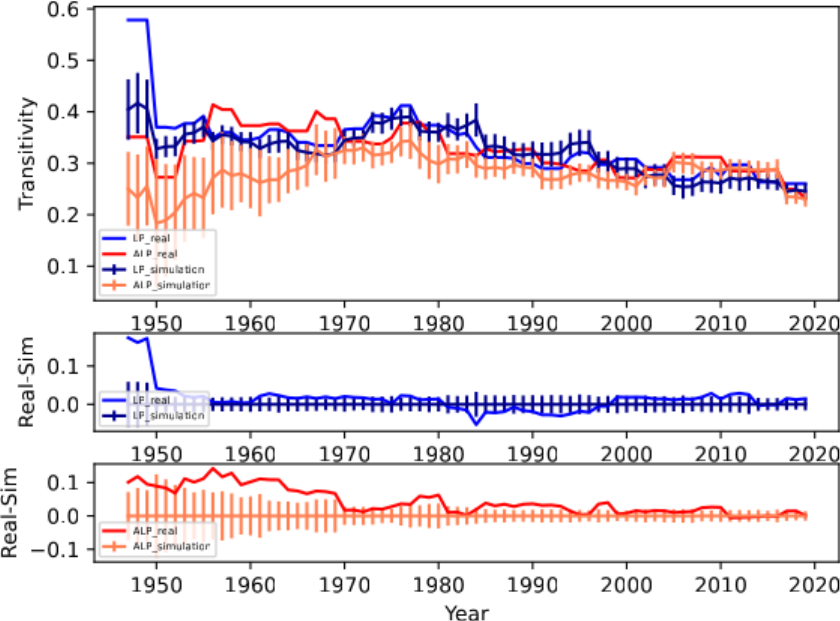}
     \caption{s=0, \textit{shuffle}}
  \end{subfigure}
\hfill
  \begin{subfigure}{0.45\textwidth}
     \includegraphics[width=\textwidth]{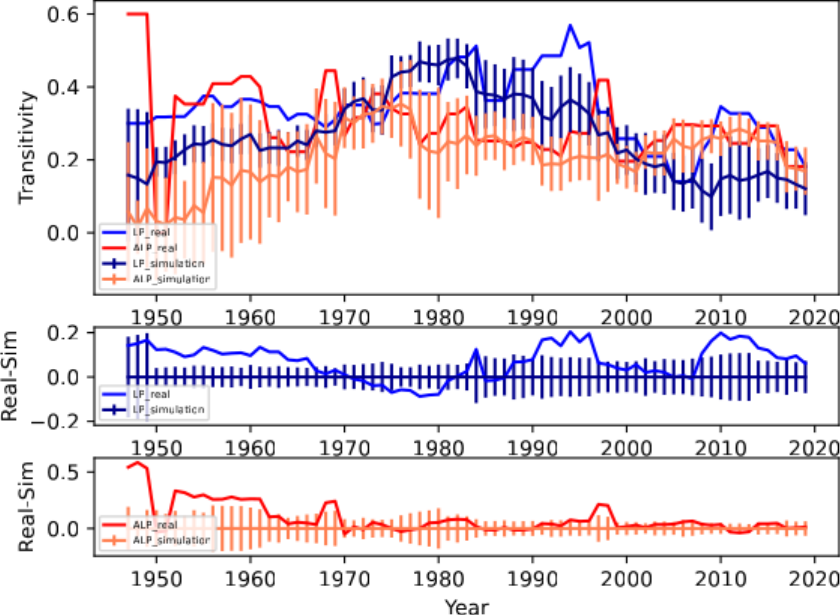}
     \caption{s=2, \textit{shuffle}}
    \end{subfigure}
    \caption{The diagrams (a) - (b) show the comparison between transitivity of s-line graphs with s being 0 and 2 respectively for real and \textit{free choice} simulated data. The diagrams (c) - (d) show the same comparison with simulation \textit{shuffle}.}
\label{classtra}
\end{figure}
The attribute networks are s-line graphs derived from the attribute hypergraphs. They have attributes as nodes and MPs as edges. As shown in Figure \ref{classamf} and Figure \ref{classtra}, the overall trend for the simulation average maximal flow for both LP and ALP with s = 0 follow roughly the trend for the average degree of the networks. The transitivity and the average maximal flow when s = 2, however, exhibit an opposite trend. We think it again could be a result of size dependency of average maximal flow and filtering the networks with higher s reduce such dependency to an extent.

In terms of the average maximal flow, we observe that the trends seem to converge when $s > 2$. The convergence of average maximal flow with high s is not surprising as filtering can be effective in extracting major structural information and is therefore relatively stable against changes in edge weight definitions. Therefore, we conclude that we can identify major structural landscapes of the networks by observing the persistent features from s = 0 to s = 2.

In terms of the average maximal flow of ALP attribute networks, we can see immediately there are two periods (before 1970s and a few years around late 19702s) where the real data exhibits unusually high values compared with both simulation methods for the two normalised weight definitions. This was stable against increase of the s-filtration to s=2. The high average flows during these two periods were even more obvious against the \textit{shuffle} simulation. It shows that the source of this high connectivity is not the size of the networks, the number of career options available, nor the distribution of MPs in different career backgrounds. But since these two peaks were not significant in transitivity, less obvious in the unweighted networks and more obvious in the directed weighted networks, we can say part of the cause would be the strong ties (possibly asymmetrical ones) among attributes, such as edges between big nodes like employed professionals and other small ones. 

Transitivity in the ALP attribute networks shows another source for the unusual high connectivity during these two periods. As shown in Figure \ref{classtra}, these two peaks can be found in the transitivity against \textit{shuffle} simulation baseline, though a weak one around late 1970s. Accordingly, we observed bouquet structures before 1970s centering humanities background and weak bouquet structures around late 1970s. Therefore, we confidently conclude that ALP attribute networks are more structurally connected during these two periods, with strong ties and weak bouquet structures being part of the cause.

Additionally, there is a slight peak (1997-1998) for the ALP in transitivity with s = 2 for both simulation methods, which is also observed in average maximal flow when $s = 2$. This is due to some satellite structures. Similar to the peak in LP attribute networks when s = 2 around 1990s. This is merely a result of disconnected networks after high  s filtration.

The Liberal Party, in terms of the average maximal flow, before 1980s, the consensus from almost all plots is that the LP attribute networks were more connected than expected from simulated baselines. The fact that the real AMF was slightly lower than the simulation for s=0, but became significantly higher than the simulation for s=2, shows that this additional connection is partly caused by the strong ties in the networks. Different from the ALP attribute networks, the higher connectivity of the LP attributes networks during this period is lessened against the \textit{shuffle} simulations. While the \textit{shuffle} does not completely bridge the gap between simulations and real data, the significant decrease in the disagreement shows that the distribution of MPs in different career choices is very important to the higher connectivity observed, for example, the high number of MPs who were employed professionals. Transitivity for LP attribute networks before 1980s also supports this idea of higher connectivity. Within the period before 1980s, it is visible that there are two peaks, one around 1955 and another one around 1975. Since the one around 1975 seems to be more prominent in the directed networks, the strong ties implied during that time are likely to be asymmetrical as well between large attributes such as employed professionals and others.

In summary, we observe that on average both ALP and LP attribute networks are more connected than simulations, though for very different reasons. Interestingly, both parties' background attribute networks converge to simulation around 1980, corroborating studies that find evidence of party recruitment convergence in Australia \cite{miragliotta_legislative_2012}.

\section{Conclusion}

This paper has introduced a novel approach to the conceptualisation and analysis of politician's background data. Our attributes-as-networks approach identifies and addresses several important challenges to leveraging network information from politicians' backgrounds. This contribution demonstrates significant potential for researchers of legislative politics since the collection of contemporary and historical background data of political elites is growing rapidly \cite{gobel_comparative_2021}.

Our central findings confirm the utility of our approach, as we demonstrate the convergence in party political recruitment. The low connectedness of trade unionist and manual laborer principal occupations, relative to elite occupations contributes to explanations of the decline of working-class representation within social democratic parties, while the emergence of the highly connected senior management attribute describes the diversification in the LP away from recruitment from the professions. 

The study makes possible several opportunities for further study. First, a dynamic analysis of the relationship between attribute node centrality and outcome variables such as ministerial selection and leadership would be of further value in demonstrating the importance of networks in the explanation of political career outcomes. Second, a useful application of our method would be to expand the analysis to comparative data to examine trends in MP professional networks across political systems. Finally, we propose an examination of attribute networks in other elite political settings such as cabinets, where reshuffles and dissolutions may be related to network structures among ministers, and the judiciary, as we hypothesise that networks may determine stable voting coalitions. 

\section{Acknowledgement}
The authors thank Rob Ackland, Sam Barton, Kate Turner, Vanessa Robins, Catherine Greenhill, Jon Slapin, and Patrick Armstrong for helpful conversations and feedback for this paper.

\singlespacing

\bibliography{1}
\newpage

\onehalfspacing

\appendix
\section{Preliminary Background on graphs and hypergraphs \label{prelim}}

In this section, we give a preliminary introduction to graph and hypergraph theory along with our notation. A weighted undirected graph is an ordered triple G = (V, E, $\phi$), where V is the set of vertices or nodes, E is the set of edges and $\phi : E \rightarrow R$, a function that maps each edge to its weight. Each weighted undirected graph with n nodes can be stored as a $n \times n$ adjacency matrix $A(G)=[a_{ij}]$ with entries defined as: 
\begin{equation}
  a_{ij} =
    \begin{cases}
      w_{ij} & \text{if $e_{ij} \in E$}\\
      0 & \text{if $e_{ij} \not\in E$}
    \end{cases}       
\end{equation}

In an unweighted undirected graph, all $w_{ij}$ would be set to 1. The degree of nodes in weighted or unweighted undirected graphs is defined to be the number of edges the node is contained in. If the graph is directed, instead of taking the unordered pair in E to R, $\phi$ takes ordered pairs of E to R. As a result, the corresponding adjacency matrix can be asymmetric.

An undirected hypergraph H = (V, E) is a generalisation of the graph where each hyperedge $e \in E$ can join any number of vertices (see Fig \ref{setup}), in contrast to only two in ordinary graphs. The cardinality or size of a hyperedge is the number of vertices that lie in the hyperedge, and the degree of a vertex counts the number of hyperedges it is incident with. Each undirected hypergraph H with n nodes and m hyperedges can be represented using an $n \times m$ incidence matrix $I(G)=[b_{ij}]$ with entries:
\begin{equation}
  b_{ij} =
    \begin{cases}
      1 & \text{if $v_i \in e_j$}\\
      0 & \text{if $v_i \not\in e_j$}
    \end{cases}       
\end{equation}

From this representation, one can easily generate the number of vertices in each hyperedge and the number of hyperedges with which each vertex coincides. Direct analysis of hypergraphs is challenging. Therefore, we follow the method in \cite{barton2022hypergraphs} and collapse the hypergraphs to s-line graphs for further analysis. For s, a positive integer, an s-line graph $L_H$ = ($V'$,$E'$) from a hypergraph H is defined as:
\begin{align}
V' = E(H) ~~ \text{and} ~~ {e_i,e_j} \in E' \iff |e_i \cap e_j| \geq s
\end{align}
where E(H) is the edge sets of H.

In other words, the s-line graph contains vertices that represent the original hyperedges and these vertices are joined by edges where the hyperedges overlap by $s$ or more. The resulting s-line graphs are no longer hypergraphs, and can therefore be analysed as ordinary networks.

\section{Degree and centrality distribution}

\begin{figure}
  \begin{subfigure}{0.24\textwidth}
     \includegraphics[width=\textwidth]{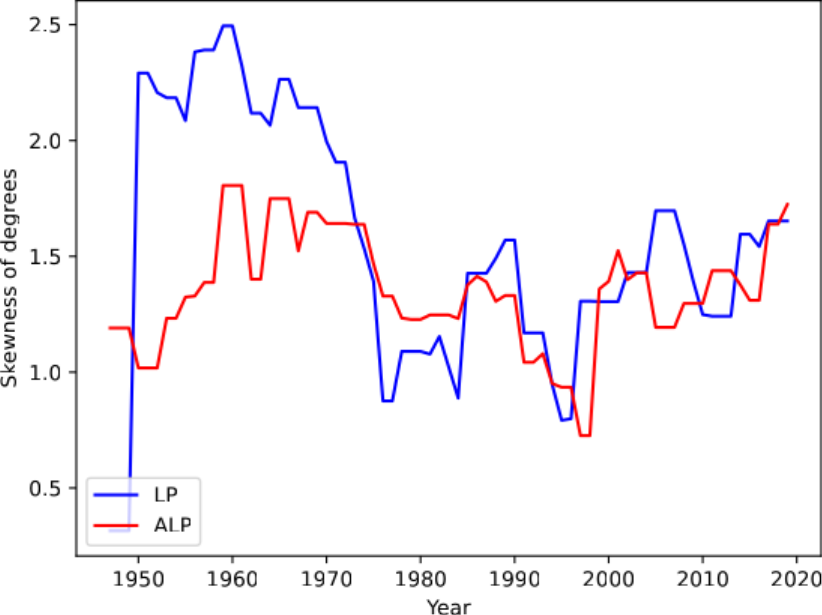}
     \caption{}
  \end{subfigure}
    \hfill
  \begin{subfigure}{0.24\textwidth}
     \includegraphics[width=\textwidth]{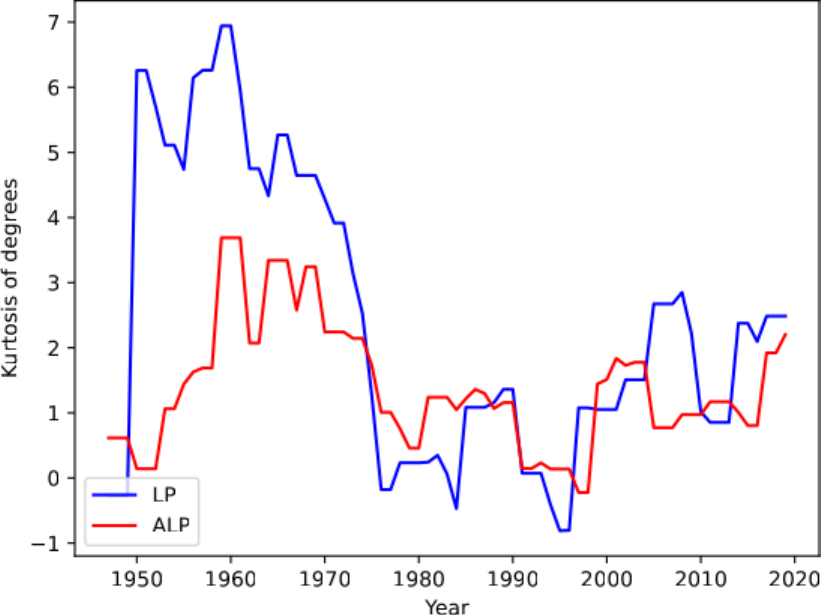}
     \caption{}
  \end{subfigure}
  \hfill
    \begin{subfigure}{0.24\textwidth}
     \includegraphics[width=\textwidth]{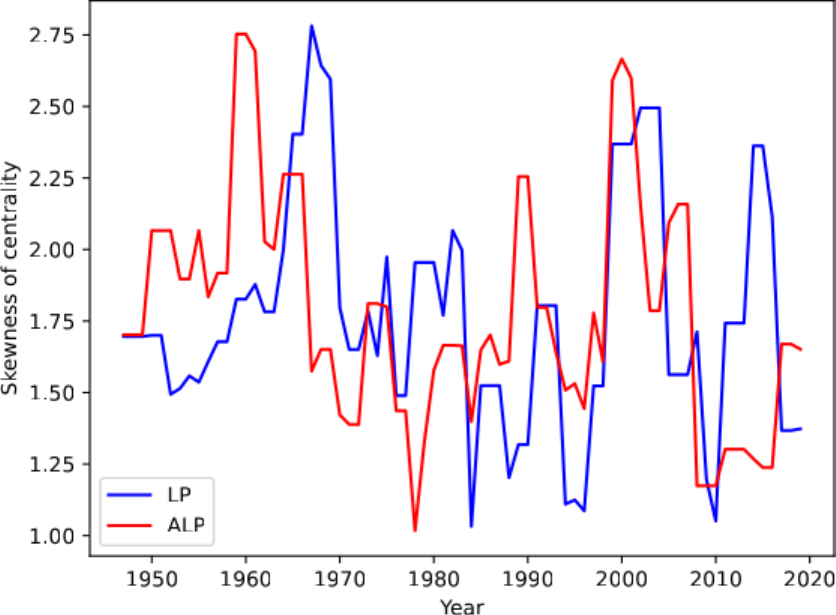}
     \caption{}
  \end{subfigure}
    \hfill
  \begin{subfigure}{0.24\textwidth}
     \includegraphics[width=\textwidth]{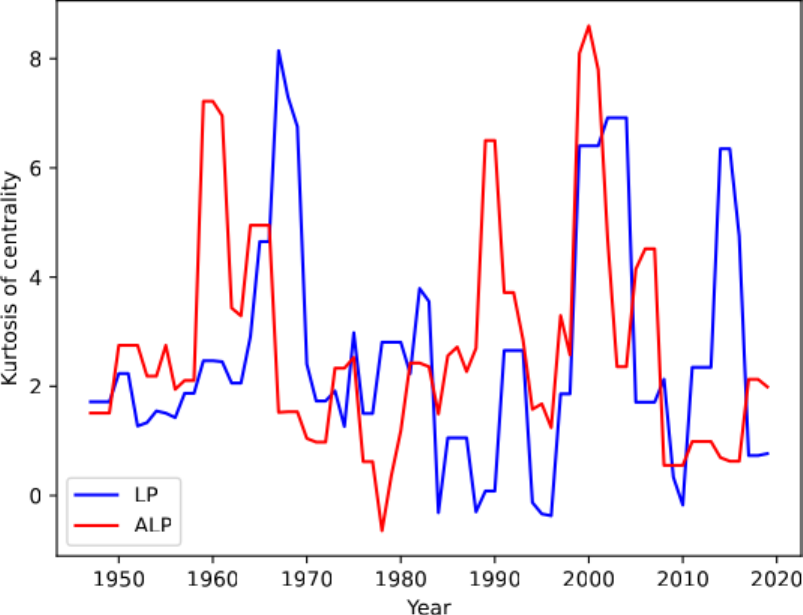}
     \caption{}
  \end{subfigure}
  \hfill
    \begin{subfigure}{0.24\textwidth}
     \includegraphics[width=\textwidth]{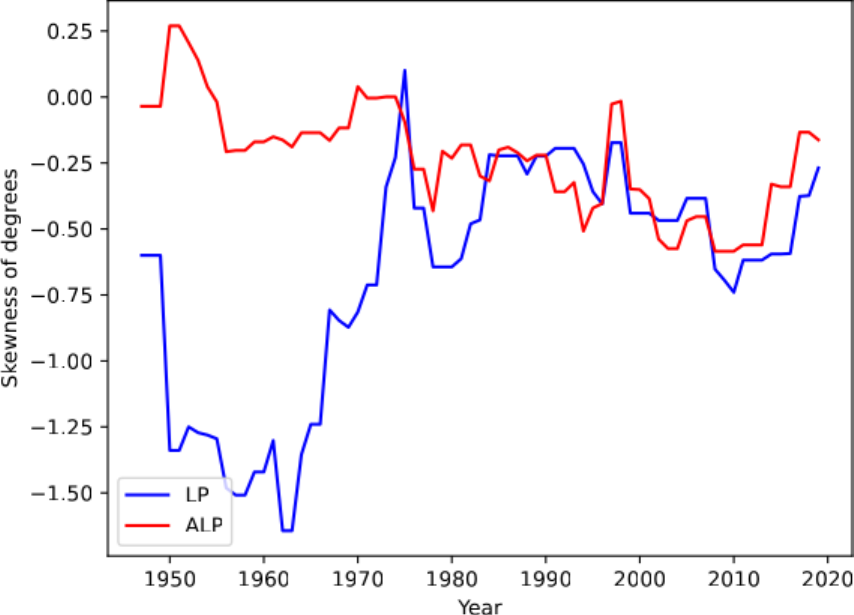}
     \caption{}
  \end{subfigure}
    \hfill
  \begin{subfigure}{0.24\textwidth}
     \includegraphics[width=\textwidth]{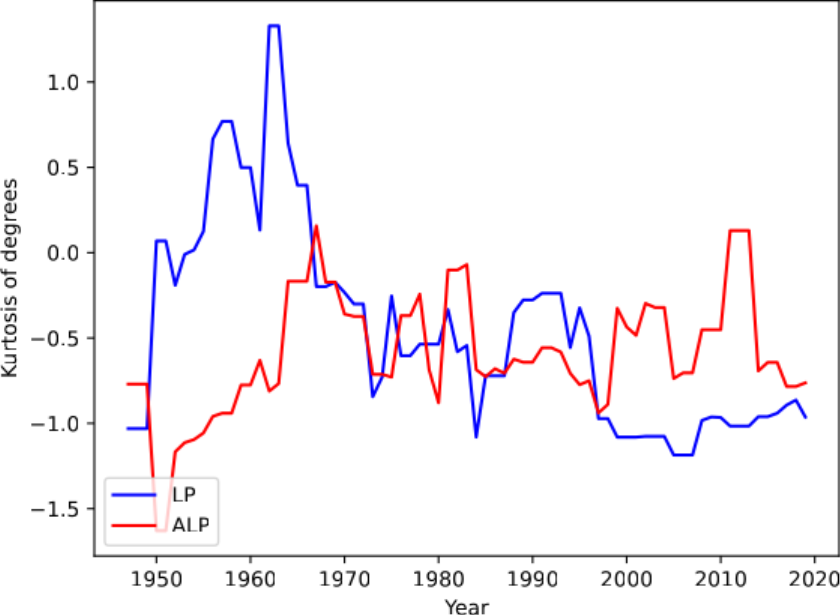}
     \caption{}
  \end{subfigure}
  \hfill
    \begin{subfigure}{0.24\textwidth}
     \includegraphics[width=\textwidth]{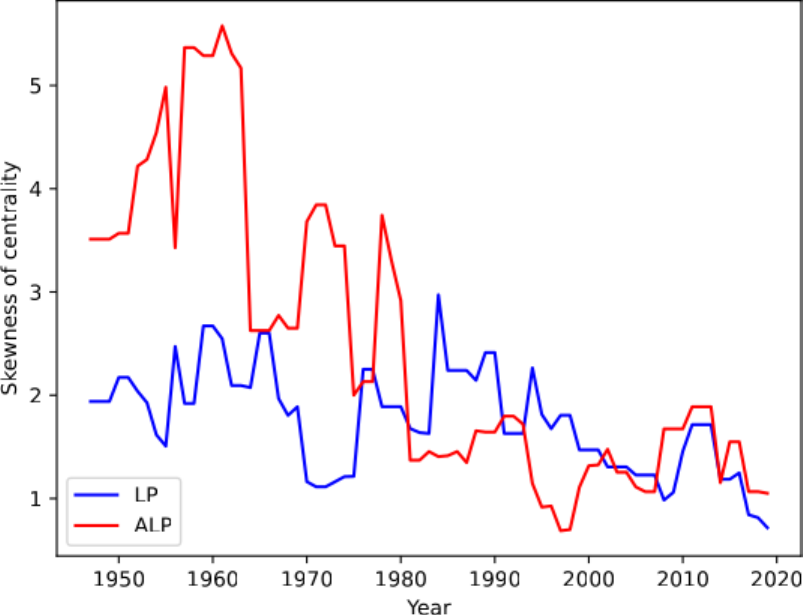}
     \caption{}
  \end{subfigure}
    \hfill
  \begin{subfigure}{0.24\textwidth}
     \includegraphics[width=\textwidth]{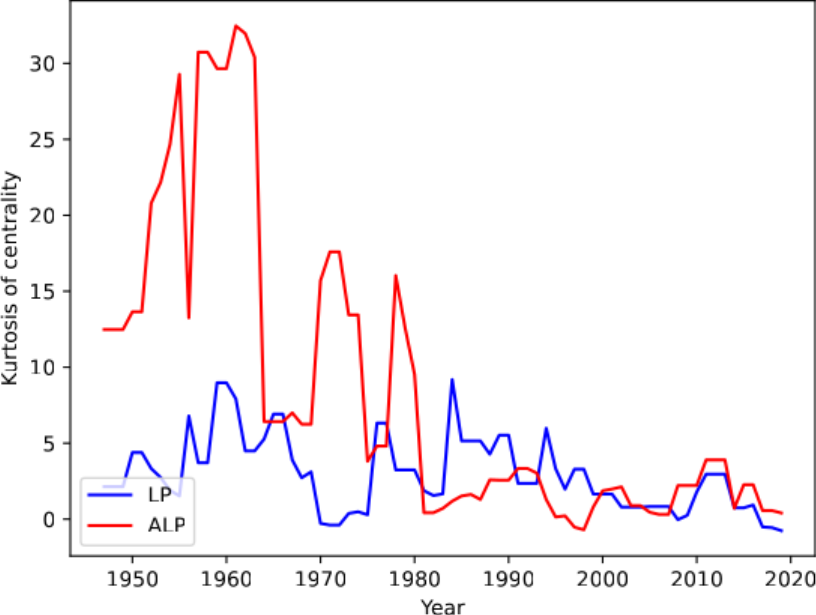}
     \caption{}
  \end{subfigure}
  \caption{(a)-(d) are for analysis of the classifier networks: (a) skewness and (b) kurtosis of the degree distribution and (c) skewness and (d) kurtosis of the centrality distribution. (e)-(h) are for analysis of the MP networks: (e) skewness and (f) kurtosis of the degree distribution and (g) skewness and (h) kurtosis of the centrality distribution.}
  \label{skew}
\end{figure}

\begin{figure}
  \begin{subfigure}{0.24\textwidth}
     \includegraphics[width=\textwidth]{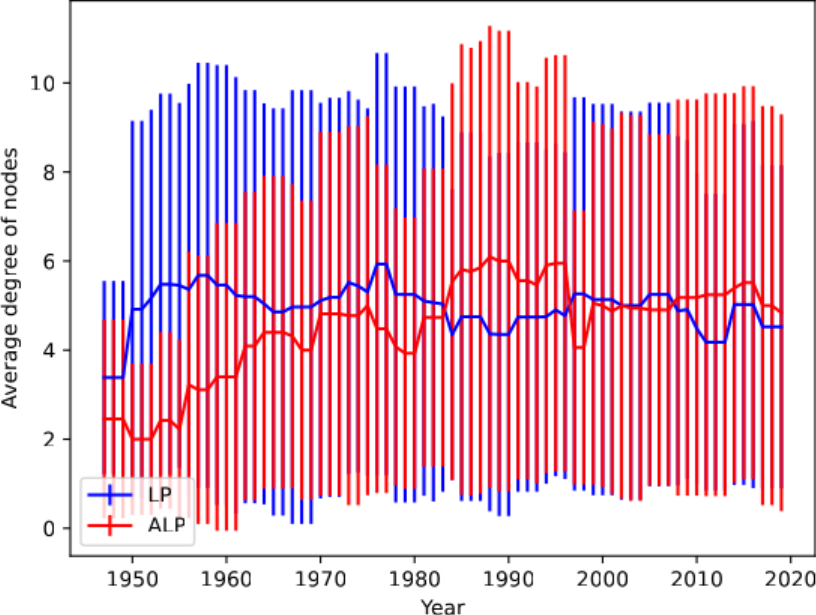}
     \caption{}
  \end{subfigure}
\hfill
  \begin{subfigure}{0.24\textwidth}
     \includegraphics[width=\textwidth]{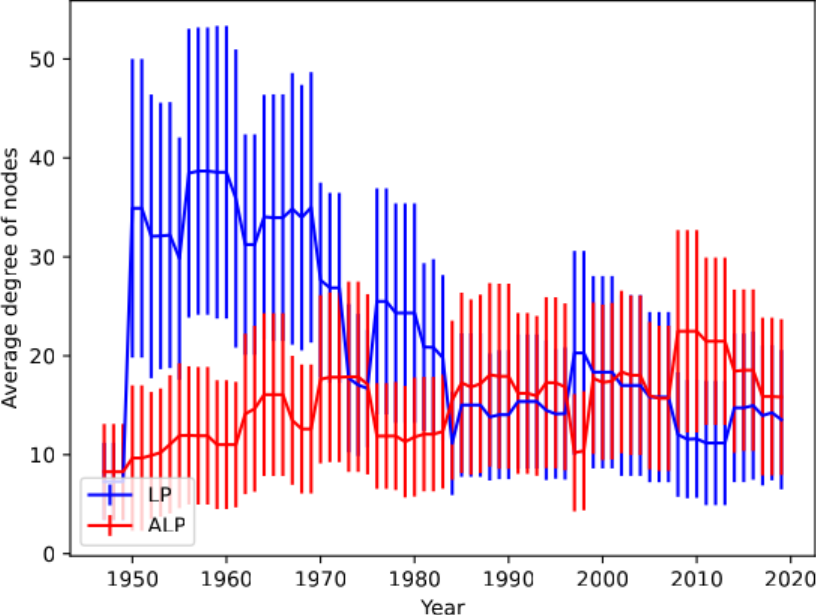}
     \caption{}
  \end{subfigure}
  \begin{subfigure}{0.24\textwidth}
     \includegraphics[width=\textwidth]{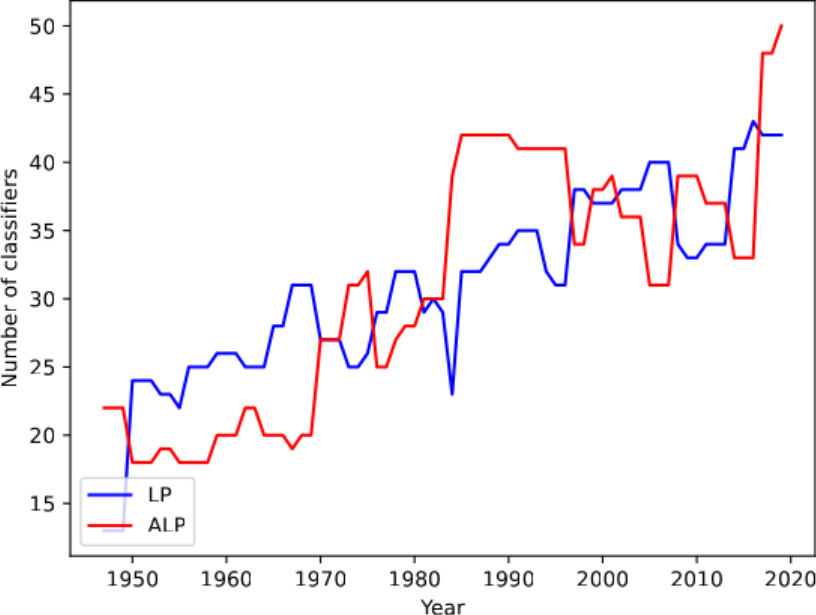}
     \caption{}
  \end{subfigure}
\hfill
  \begin{subfigure}{0.240\textwidth}
     \includegraphics[width=\textwidth]{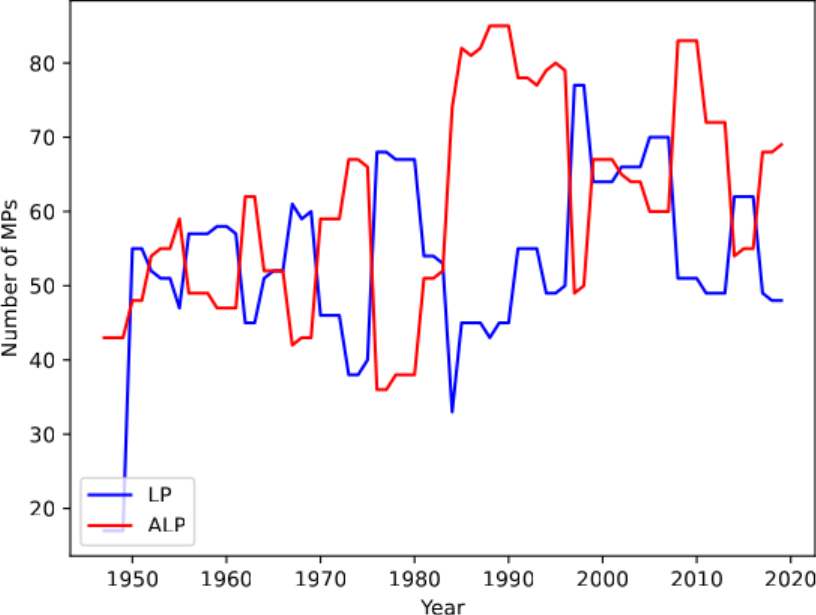}
     \caption{}
  \end{subfigure}
    \caption{The diagrams (a) and (b) show the average degrees for nodes in each year for (a) the classifier networks and (b) the MP networks. The diagrams (c) and (d) show the number of nodes in each year for (c) the classifier networks and (d) the MP networks.}
    \label{number}
\end{figure}

We also use the centrality distribution as supplementary information (See Figure \ref{skew} and \ref{number}). There are many flavours of centrality measures, the one we implemented is betweenness centrality. The definition for the betweenness centrality is as follows:
\begin{align}
c_B(v) = \sum_{s,t \in V} \frac{\sigma(s,t|v)}{\sigma(s,t)}
\end{align}
where V is the set of nodes, $\sigma(s,t)$ is the total number of shortest paths from s to t, and $\sigma(s,t|v)$ is the number of those paths that contains v while v is not s nor t. If v = s or v = t or s = t, the centrality is defined as 0 \cite{brandes2008variants}. The weighted undirected networks were used for calculations for centrality, but the path lengths are defined as the inverse of the edge weights.

\section{The bouquet structure}
 \begin{figure}
  \begin{subfigure}{0.4\textwidth}
     \includegraphics[width=\textwidth]{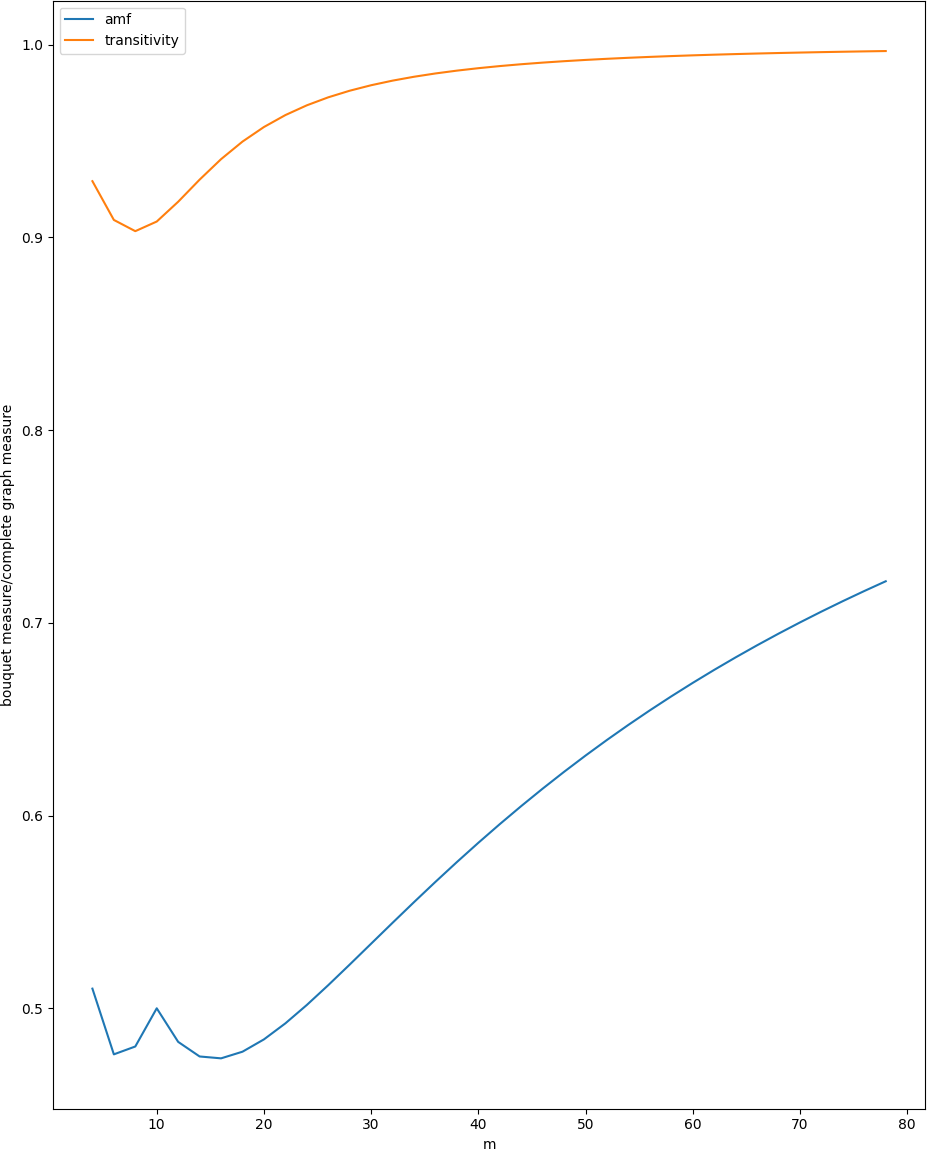}
     \caption{}
  \end{subfigure}
\hfill
 \begin{subfigure}{0.28\textwidth}
     \includegraphics[width=\textwidth]{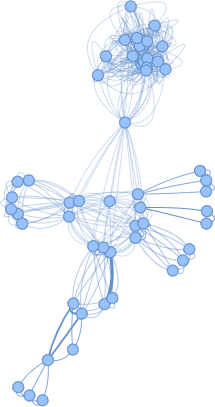}
     \caption{}
  \end{subfigure} 
\caption{(a) Comparison of the effects of the average maximal flow and transitivity in bouquet structures. We set m,n to be the number of nodes on each side of the bouquet, and n is set to 10, while m varies on the x-axis. (b) An example of the bouquet structure: the ALP MP network at year 1955.}
\label{amfvstran}
\end{figure} 
The bouquet structure refers to structures that have a small community (the most drastic version being the small community as one node) that connects two much larger closely connected communities (close to complete subnetworks). The justification for the effects on this structure on the two graph measures (average maximal flow and transitivity) here looks at structures with one node as the connecting bridge between two complete subnetworks with size n and m respectively. When n = m, it is not hard to show that the average maximal flow would be exactly half of the average maximal flow of the corresponding complete network of size n+m+1. Because for each pair of nodes in the bouquet structure with equal ends with size n, the number of paths between them is constant at n. While in the complete graph of size 2n+1, the number of paths between each pair of nodes is constant at 2n. \\

As we increase the size of one side of the bouquet, both the average maximal flow and the transitivity will grow since the network is getting more and more connected. As we see in figure \ref{amfvstran}(a), transitivity is very much unaffected by this structure and approaches to 1 very quickly. Average maximal flow, however, only starts to pick up when one of the end starts to be get past more than four times the size of the other end.\\

Moreover, such structures can also show up on the skewness of centrality distribution sometimes, as the center node of the bouquet is expected to have much higher centrality compared to the rest.

\section{Satellite}
\begin{figure}
    \centering
    \includegraphics[width=0.7\textwidth]{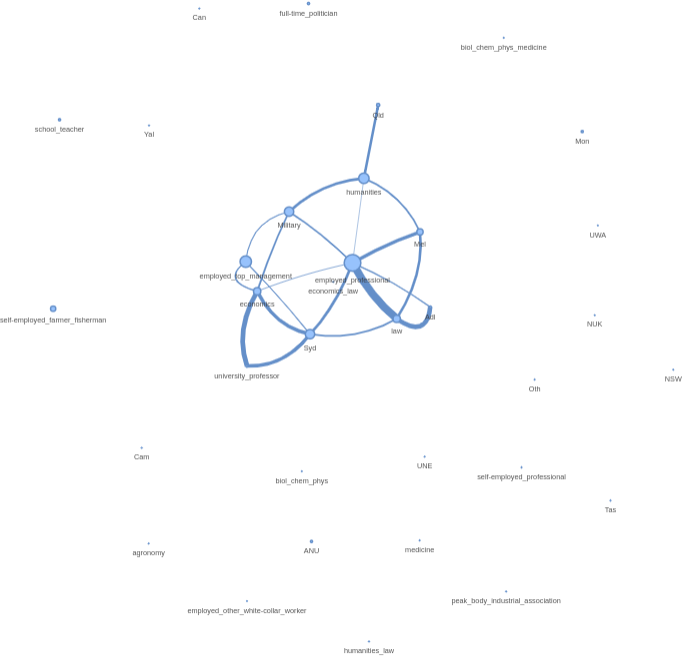}
    \caption{An example of satellite structure: LP classifier network at year 1990 with s = 2.}
        \label{satellite}
\end{figure}

By definition of transitivity and average maximal flow, this structure will lead to higher transitivity and low average maximal flow (see Fig \ref{satellite}).

\section{Asymmetry of hyperedge crossing \ref{substruct}(c)}
%  \begin{figure}
%  \centering
%    \includegraphics[width=0.5\textwidth]{LP_1974_0.png}
% \caption{Example of a network with many strong edges between large and medium hyperedges and very small hyperedges. This is the classifier network of LP from 1974 with s=0.}
% \end{figure} 
Here we show that how asymmetry affects the average maximal flow of the directed weighted networks compared to the undirected weighted networks according to the definitions of the two types of networks in section \ref{MCdef}.\\

The way we show it is to compare the changes in the average maximal flow from symmetrical crossings to asymmetrical crossing for the two sets of networks, while fixing the overlap. We set the number of nodes in each hyperedge as A and B respectively, and the number of overlap nodes in the intersection of two hyperedges as n (see figure \ref{substruct}(c).\\

We start with the undirected weighted networks. For the symmetrical case, where A = B, the average maximal flow is:
\begin{align}
\text{amf}_{\text{sym}} = \frac{n}{2B - n} 
\end{align}
Then we change the size of one hyperedge from A to A', and now the average maximal flow is:
\begin{align}
    \text{amf}_{\text{asym}} = \frac{n}{A'+B-n}
\end{align}
Now we compute the difference between those two. We set $n = r'\times B$ and $A' = r \times B$, and we know that $r > r'$, $r' < 1$:
\begin{align}
  D &= \frac{n}{2B - n} -  \frac{n}{A'+B-n} \\
  &= r' \frac{1-r}{(r+1-r')(2-r')}
\end{align}
Therefore, the relative difference is:
\begin{align}
    \frac{D}{\text{amf}_{\text{sym}}} = \frac{1-r}{r+1-r'}
\end{align}

Similarly with the directed weighted networks, for the symmetrical case, the average maximal flow is:
\begin{align}
\text{amf'}_{\text{sym}} = \frac{n}{B} 
\end{align}
And
\begin{align}
    \text{amf'}_{\text{asym}} = \frac{1}{2}(\frac{n}{A'} + \frac{n}{B})
\end{align}
The difference is:
\begin{align}
   D' &= \frac{1}{2}(\frac{n}{A'} + \frac{n}{B}) - \frac{n}{B} \\
   &= \frac{1}{2}r' \frac{1-r}{r} 
\end{align}
And the relative difference is:
\begin{align}
    \frac{D'}{\text{amf'}_{\text{sym}}} = \frac{1-r}{2r}
\end{align}
As $r \rightarrow 0$ (as A decreses in size), $r' \rightarrow r$, we can see that the relative difference in the directed option is going be a lot higher than that of the undirected option. When $r \rightarrow$ a very large number, the undirected relative difference goes towards -1, while the directed relative difference goes towards -1/2.  
\section{Increase/decrease in overlap}
\begin{figure}
    \centering
    \includegraphics[width=0.6\textwidth]{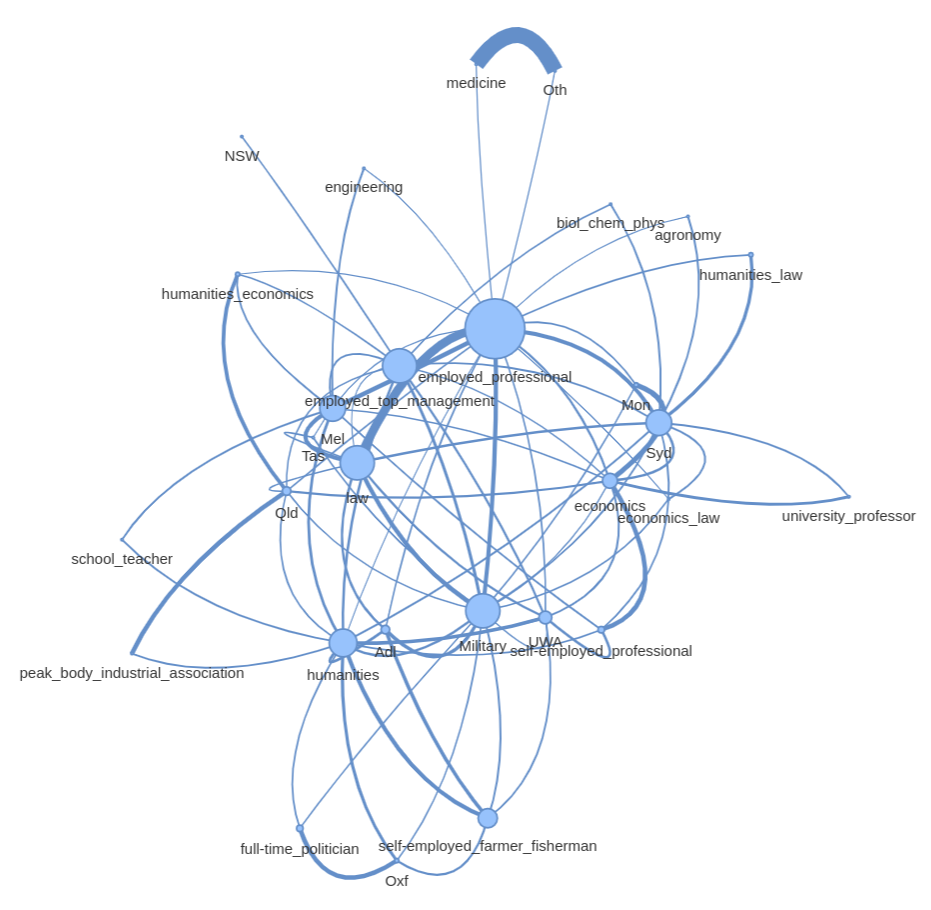}
    \caption{Example of this phenomenon potentially explaining why the average maximal flow for the undirected weighted network was very low for classifier network for LP from 1980 with s=0.}
\end{figure}
We show here that the effects on the average maximal flow with changes in the size of the intersection between hyperedges are more dramatic in undirected weighted networks than the directed weighted networks. \\

We again show the effects in a simple system where two hyperedges intersect. We again set the sizes of two hyperedges to be A and B, and the starting intersection as n. Then we change n to n'.\\

The percentage changes in the directed weighted networks is
\begin{align}
    \frac{amf_{before}}{amf_{after}} = \frac{n}{n'}
\end{align}
The percentage change in the undirected weighted networks is:
\begin{align}
    \frac{amf_{before}}{amf_{after}}= \frac{n}{n'} \times \frac{A+B-n'}{A+B-n}
\end{align}
It can be shown that no matter $n'>n$ or $n'<n$ the difference in the undirected weighted network would always be larger than the directed weighted network.
\section{Technical details}
Visualisation of these networks was done by Pyvis and networkx.

\end{document}